\documentclass[aip,reprint,english,a4paper]{revtex4-1}
\usepackage{setspace}
\usepackage{upgreek}
\usepackage{amsmath}
\usepackage{amsfonts}
\usepackage{amstext}
\usepackage{mathtools}
\usepackage{amssymb}
\usepackage{siunitx}
\usepackage{color}
\usepackage[normalem]{ulem}
\usepackage{float}
\usepackage{multirow}
\usepackage{graphicx}
\usepackage{subfigure}
\usepackage{bm}
\usepackage[linesnumbered,ruled,vlined]{algorithm2e}

\newcommand*\diff{\mathop{}\!\mathrm{d}}
\newcommand{\KL}[2]{D_{\text{\textsc{kl}}} \left[#1 \,\middle\Vert\, #2 \right]}

\newcommand{\SMmatern}{S1}
\newcommand{\SMinducing}{S2}
\newcommand{\SMvae}{S3}
\newcommand{\SMdendrograms}{S4}
\newcommand{\SMlysozymeinducing}{S5}
\newcommand{\SMlysozymemosaicmatrix}{S6}
\newcommand{\SMlysozymejawstructure}{S7}
\newcommand{\SMlysozymehistograms}{S8}
\newcommand{\SMlysozymeembedding}{S9}

\newcommand{\SMsecMatern}{I}
\newcommand{\SMsecInducing}{II}
\newcommand{\SMsecParameters}{III}

\newcommand{\SMtabhyperparameters}{S1}
\newcommand{\SMtabmosaic}{S2}

\begin{document}
\singlespacing
\author{Georg Diez}
\email{georg.diez@physik.uni-freiburg.de}
\affiliation{Biomolecular Dynamics, Institute of Physics, University of Freiburg, 79104 Freiburg, Germany}
\author{Nele Dethloff}
\affiliation{Biomolecular Dynamics, Institute of Physics, University of Freiburg, 79104 Freiburg, Germany}
\author{Gerhard Stock}
\email{stock@physik.uni-freiburg.de}
\affiliation{Biomolecular Dynamics, Institute of Physics, University of Freiburg, 79104 Freiburg, Germany}
\email{stock@physik.uni-freiburg.de}

\date{\today}

\title{Recovering Hidden Degrees of Freedom Using Gaussian Processes}
\date{\today}
\begin{abstract}
    Dimensionality reduction represents a crucial step in extracting meaningful 
    insights from Molecular Dynamics (MD) simulations.
    Conventional approaches, including linear methods such as 
    principal component analysis as well as various autoencoder
    architectures, typically operate under the assumption of independent
    and identically distributed data, disregarding the 
    sequential nature of MD simulations.
    Here, we introduce a physics-informed representation learning framework
    that leverages Gaussian Processes combined with variational autoencoders
    to exploit the temporal dependencies inherent in MD data.
    Time-dependent kernel functions---such as the Matérn
    kernel---directly impose the temporal correlation structure of the 
    input coordinates onto a low-dimensional space, preserving 
    Markovianity in the reduced representation while faithfully capturing 
    the essential dynamics.
    Using a three-dimensional toy model, we demonstrate that this 
    approach can successfully identify and separate dynamically
    distinct states that are geometrically indistinguishable due to 
    hidden degrees of freedom. 
    Applying the framework to a $50\,\mu$s-long MD trajectory
    of T4 lysozyme, we uncover dynamically distinct conformational substates
    that previous analyses failed to resolve, revealing functional 
    relationships that become apparent only when temporal correlations 
    are taken into account.
    This time-aware perspective provides a promising framework
    for understanding complex biomolecular systems, in which conventional
    collective variables fail to capture the full dynamical picture.
\end{abstract}

\maketitle

\section{Introduction}
Biomolecular systems such as proteins are inherently dynamic, such that functional
biomolecular processes are governed by structural rearrangements and
conformational changes.
Molecular dynamics (MD) simulations have become a cornerstone
for studying these dynamics, but the resulting high-dimensional datasets easily
obscure the desired physical insights.\cite{Berendsen07}
This motivates the development of dimensionality reduction strategies that
simplify the complex 3$N-$dimensional trajectories into interpretable
collective variables that reveal key mechanisms and structural transitions 
of the protein.\cite{Bolhuis00,McGibbon17,Sittel18,Fleetwood21,Diez22,Wang21,Glielmo21}
Classical and well-established approaches such as principal component analysis
and time-lagged independent component analysis have proven useful
for linear feature extraction, focusing on maximizing either variance or
time scales, respectively.\cite{Sittel17, Perez-Hernandez13} 
More recently, nonlinear representation learning techniques have emerged 
that offer greater flexibility in capturing the complex features of 
proteins.\cite{Glielmo21}
Some of these variants explicitly incorporate temporal 
information,\cite{Rohrdanz13, Mardt18, Chen19} enabling powerful, 
data-driven extraction of low-dimensional
representations.\cite{Lemke19,Lemke19b,Wang21,Belkacemi21}
Such neural networks leverage their capability as universal function
approximators\cite{Hornik89} to encode high-dimensional data into a reduced
latent space and then reconstruct it, thus capturing the most essential 
features while filtering out less relevant details.

Among these, Variational Autoencoders (VAEs) and their extensions have shown 
promise in learning latent spaces that informatively encode the essential 
features of biomolecular motions on the one hand while enabling 
generative modeling by creating unseen samples on the other hand.
\cite{Kingma13, Doersch16, Ribeiro18, Varolgunecs20, Tian21}
Recent adaptations of VAEs have further enhanced their applicability to 
MD simulation data through more specialized priors that account for 
the complex free energy landscapes of proteins.\cite{Wang21}
For example, the VampPrior (Variational Mixture of Posteriors) 
technique extends the standard VAE framework by
implementing a more flexible prior, consisting of a mixture
of variational posteriors conditioned on learnable pseudo-inputs.\cite{Tomczak18}
Such pseudo-inputs can be thought of as characteristic frames in the 
trajectory that represent core patterns in the data, such as e.g. 
metastable conformational states. 
Incorporating these directly into the prior,
allows for better matching the true latent space representation of 
the complex biomolecular conformational space.\cite{Ribeiro18, Wang21}
Similarly, in the work by Varolgüneş \textit{et al.}, the multi-basin
free-energy landscape is naturally captured by a Gaussian mixture 
variational autoencoder, which simultaneously performs dimensionality
reduction and clustering within a single unified framework.\cite{Varolgunecs20}

More recently, VAEs have been extended to model 
sequential data explicitly,\cite{Chung15,Girin20,Hasan21} 
addressing a common fundamental limitation of many of the above 
mentioned approaches:
the assumption that data points are independent and identically 
distributed (i.i.d). 
While this assumption is mathematically convenient
(e.g. it facilitates batch-wise network training), it fails to
capture the temporal dependencies in MD simulations. 
Thus, it may not account for the time evolution of the system,
where each conformation directly depends on previous states.
Among physics-informed approaches, a particularly promising and 
interesting step is the dynamics-constrained representation learning 
framework introduced by Tiwary and coworkers,\cite{Wang24} which 
restricts the latent representation to follow overdamped Langevin dynamics.
Incorporating Langevin dynamics directly into the prior establishes a
feature extraction framework that is driven by statistical mechanics 
rather than user-made assumptions about the underlying distribution 
of the collective variables.
%
%
Another established approach for analyzing biomolecular dynamics 
are hidden Markov models, which share our goal of
unveiling the underlying Markovian dynamics from observable 
data.\cite{Rabiner86,Noe13}
While these models describe transitions between unobserved metastable states
using a discrete state representation, our approach addresses the 
same challenge through a continuous latent space representation
that preserves Markovianity.

Here, we introduce a physics-informed approach that leverages
Gaussian Processes (GPs) to model the temporal correlations in 
sequential data like MD simulations and therefore directly address 
the i.i.d.\ assumption.
Unlike principal component analysis or traditional autoencoder 
architectures that treat each frame independently, the proposed 
GP-VAE framework explicitly encodes time-dependent relationships 
between data points through kernel functions. 
Motivated by its relationship to stochastic processes that 
underlie Brownian motion, we employ the Matérn kernel,
which enables us to naturally encode 
Markovianity and temporal correlations in the latent space
representations.\cite{Rasmussen03}
This facilitates both dimensionality reduction as well as 
discovery of physically relevant and kinetically faithful representations 
that reflect hidden or unobserved substates.
By building upon temporal relationships between the data points,
our approach can separate dynamically distinct states that are 
otherwise geometrically indistinguishable due to 
hidden or unobserved degrees of freedom.

\section{Theory}
\subsection{Bayesian Terminology: Prior, Likelihood and Posterior}
When studying protein dynamics, each snapshot of a protein structure $\bm{x}$
resides in a very high-dimensional conformational space 
(e.g. $3 N$ atomic coordinates).
Yet, most crucial structural rearrangements, such as the formation and
breaking of secondary structures by local contacts 
can often be described by far fewer degrees of freedom.\cite{Sittel18,Noe20}
This observation motivates the introduction of hidden latent variables,
which represent low-dimensional hidden descriptors $\bm{z}$ of 
the protein dynamics (commonly referred to as collective variables).
These latent variables capture the most important features that are used to 
describe the protein's conformational changes, while simultaneously 
discarding irrelevant details such as e.g. thermal fluctuations.

To formalize this idea, we consider a latent variable model in which each
observed conformation $\bm{x}$ in the full space is associated with an 
unobserved latent variable $\bm{z}$. 
The generative process can probabilistically be described as follows:
\begin{align*}
    \bm{z} &\sim p(\bm{z}),\\
    \bm{x} &\sim p(\bm{x}|\bm{z}),
\end{align*}
where $\sim$ means "is sampled from", 
$\bm{z} \in \mathbb{R}^d$, $\bm{x}\in \mathbb{R}^D$ and $d \ll D$
reflects the task of dimensionality reduction.
Furthermore, $p(\bm{z})$ is the prior distribution that reflects our 
assumptions about the latent space without having observed any data yet.
Commonly, a simple Gaussian prior 
$p(\bm{z})=\mathcal{N}(\bm{z}| \bm{0}, \bm{I})$ is assumed.\cite{Kingma13, Doersch16}
The likelihood $p(\bm{x}|\bm{z})$ encodes how a specific latent
variable $\bm{z}$ leads to an observation $\bm{x}$, 
hence acting as a "recipe" for reconstructing the 
protein's full conformation $\bm{x}$ given the collective 
variable $\bm{z}$.
In this generative framework, each protein conformation $\bm{x}$ is 
viewed as arising from a latent collective variable $\bm{z}$ via
\begin{align}
    p(\bm{x}, \bm{z}) = p( \bm{x}|\bm{z}) p(\bm{z}).
    \label{eq:VAE_gen}
\end{align}

In practice, we only observe the proteins' conformation $\bm{x}$ but 
have no direct access to the latent variables $\bm{z}$.
To evaluate how well our model explains the observed data, and thus how
well we have chosen our collective variables, we want to compute the 
marginal likelihood $p(\bm{x})$, which considers all possible latent 
variables that could possibly have generated $\bm{x}$. 
This step requires integrating over $\bm{z}$:
\begin{align}
    p(\bm{x}) = \int \diff \bm{z} \, p(\bm{x}|\bm{z}) p(\bm{z})
    \label{eq:px}
\end{align}
However, computing this integral generally becomes intractable for high 
dimensional data obtained from MD simulation due to the high dimensionality 
of both $\bm{x}$ and $\bm{z}$.

To address this challenge, instead of directly computing $p(\bm{x})$
in Eq.~\eqref{eq:px}, VAEs employ variational inference to 
introduce an approximate posterior distribution $q_\phi (\bm{z}|\bm{x})$, 
which describes how likely a particular latent variable 
$\bm{z}$ is given an observation $\bm{x}$.\cite{Kingma13, Doersch16}
This approximate posterior is parameterized by a neural network
with parameters $\phi$, allowing us to derive a tractable
lower bound on the log-likelihood.
During training, the so-called Evidence Lower Bound (ELBO)\cite{Higgins17},
which is expressed through the  loss function $\mathcal{L}_{\theta,\phi}$
is maximized:
\begin{equation}
\begin{aligned}
    \ln p(\bm{x}) \geq \mathcal{L}_{\theta, \phi} &= \underbrace{\mathbb{E}_{\bm{z}\sim q_\phi(\bm{z}|\bm{x})} \left[ \ln p_\theta (\bm{x}|\bm{z}) \right]}_\text{Reconstruction} \\ 
    &\phantom{=}- \beta \underbrace{D_\textsc{kl} \left[ q_\phi (\bm{z}|\bm{x}) || p (\bm{z})\right]}_\text{Regularization},
    \label{eq:VAE_ELBO}
\end{aligned}
\end{equation}
where $D_\textsc{kl}[\cdot||\cdot]$ is the Kullback-Leibler (KL) divergence, which 
measures how far the learned approximate posterior $q_\phi(\bm{z}|\bm{x})$
diverges from the prior $p(\bm{z})$.
Both $\theta$ and $\phi$ are learnable parameters obtained through the 
neural network training process, representing the parameters of the decoder
and encoder respectively.

This is where similarities with traditional autoencoder/information bottleneck
architectures become apparent: on the one hand, the first term encourages 
the stochastic decoder $p_\theta(\bm{x}|\bm{z})$ to faithfully
reconstruct $\bm{x}$ from latent samples $\bm{z}$ which are drawn 
according to the encoder $q_\phi(\bm{x}|\bm{z})$.
On the other hand, the regularization term $\beta D_\textsc{kl}$ prevents
the learned latent representation from drifting too far away from the assumed
prior $p(\bm{z})$, with the hyperparameter $\beta$ being an adjustable
weighting factor.\cite{Higgins17}

\subsection{Time-Aware Representations with Gaussian Processes}

In this VAE-framework, we can introduce temporal correlations 
by modifying the prior $p(\bm{z})$ such that it is conditioned on the time $t$
through GPs, yielding $p(\bm{z}|t)$.\cite{Casale18, Jazbec21, Tian24}
The rationale for using a GP as a prior over the latent space is 
straightforward:
frames close in time are considered similar---and their latent
representations $\bm{z}$ are therefore placed close to each other---while
temporally distant ones exhibit weaker correlations.
This shifts the focus from correlation among features to relationship 
among the data points themselves,\cite{Rasmussen03, Titsias09, Hensman13, Deringer21}
facilitating the utilization of the already existing time information
in MD simulations.

We define a GP by its mean function $\mu(\cdot)$ and covariance
kernel function $k(\cdot, \cdot)$.
Here, we will focus on time-dependent kernels $k(t, t^\prime)$, 
which are well-suited for MD simulations where each data point
$\bm{x}_i$ is associated with a time $t_i$.
However, GPs can generally be applied to a variety of domains, such as e.g.
view angles of objects in images,\cite{Casale18} spatial coordinates in
transcriptomics studies,\cite{Tian24} or in the automated discovery of 
small organic compounds.\cite{Mohr22}
Formally, we write
\begin{align*}
    z \sim \mathcal{GP}\bigl[ \mu(t), k(t, t^\prime)\bigr],
\end{align*}
meaning that the latent variable $z$ is distributed as a GP with mean 
function $\mu(t)$ and a covariance function $k(t,t^\prime)$ that encodes
the correlations over time.

A central modeling choice is the kernel function
$k(t,t^\prime)$, as it determines the structure of temporal correlation that
we impose on the latent space. 
For MD data, we want to preserve a simpler Markov-like structure in the latent
space, which is why the Matérn kernel is particularly well-suited.
Mathematically, it is closely related to the Ornstein-Uhlenbeck process,
which was originally introduced to model the velocity 
of a particle undergoing Brownian motion.\cite{Rasmussen06}

The Matérn kernel is given as
\begin{align*}
        k_{\nu,\ell}(t, t^\prime) &= \frac{2^{1-\nu}}{\Gamma(\nu)} \left( \sqrt{2\nu} \frac{|t - t^\prime|}{\ell} \right)^\nu K_\nu \left( \sqrt{2\nu} \frac{|t - t^\prime|}{\ell} \right),
\end{align*}
where $\nu$ specifies the smoothness of the kernel function, $\ell$ is the 
user-defined length/time scale, and $K_\nu$ is a modified Bessel function.\cite{Rasmussen06}
Typically, only half-integer values $\nu = 1/2+n$ 
(where $n \in \mathbb{N}$) are considered, because the covariance function
then simplifies to a product of an exponential and a polynomial of order $n$,
avoiding the computationally expensive Bessel functions.
Rasmussen identified $\nu=3/2$ and $\nu=5/2$ as the most interesting
cases for machine learning: $\nu=1/2$ results in overly rough, non-differentiable
processes, while $\nu > 5/2$ produces increasingly smooth processes which 
leads to a loss of the Markov property.\cite{Rasmussen06}
For our purposes, we found that $\nu=3/2$ provides an optimal balance
between temporal smoothness and preservation of Markovian dynamics:
\begin{align*}
    k_{\nu=3/2, \ell} (t,t^\prime) = \left(1 + \frac{\sqrt{3} |t - t^\prime|}{\ell}\right) \exp\left( - \frac{\sqrt{3} |t - t^\prime|}{\ell}\right).
\end{align*}
We discuss the effects of the smootheness parameter $\nu$ and the length
scale parameters $\ell$ in more details in the supplementary material (SM),
Sec.~\SMsecMatern~and  Fig.~\SMmatern.

\subsection{The GP-VAE Loss Function}
Employing a time-dependent kernel as a prior, 
the joint distribution for the GP-VAE now becomes
\begin{align*}
    p(\bm{x}, \bm{z}|t) = p(\bm{x}|\bm{z}) p(\bm{z}|t).
\end{align*}
This modifies the standard VAE ELBO objective function 
in Eq.~\eqref{eq:VAE_ELBO} to 
\begin{align}
     \mathcal{L} &= \mathbb{E}_{\bm{\tilde{z}}\sim \tilde{q}_{\phi}(\bm{\tilde{z}}|\bm{x})}\left[\log p_{\theta}(\bm{x}|\bm{\tilde{z}})\right] \nonumber \\
     &\phantom{=}- \beta \KL{q_{\phi}(\bm{z}|\bm{x},t)}{p(\bm{z}|t)},
     \label{eq:starting_point}
\end{align}
where we notationally distinguish between two posteriors: 
the traditional posterior VAE encoder output $\tilde{q}_{\phi}(\tilde{\bm{z}}|\bm{x}) = \mathcal{N}(\tilde{\boldsymbol{\mu}},\tilde{\boldsymbol{\sigma}}^2)$, 
with $\tilde{\boldsymbol{\mu}} = f_{\mu}[f_{\phi}(\bm{x})]$ and 
$\tilde{\boldsymbol{\sigma}}^2 = f_{\sigma} [f_{\phi}(\bm{x})]$
learned by neural networks, and the time-aware GP posterior 
$q_{\phi}(\bm{z}|\bm{x},t)$ that incorporates temporal 
correlations through the GP prior $p(\bm{z}|t)$.
Note that both posteriors depend on the same encoder parameters $\phi$, 
as the GP regularization transforms the encoder output using fixed 
kernel hyperparameters.
From here on, we use the notation $\mathbb{E}[\cdot]_{\tilde{q}_{\phi}(\bm{z}|\bm{x})}$ 
instead of $\mathbb{E}[\cdot]_{\bm{z} \sim \tilde{q}_{\phi}(\bm{z}|\bm{x})}$ 
for the sake of brevity.

Although conceptually simple, this modification introduces substantial
computational hurdles that set GP-VAEs apart from standard VAEs.
For large datasets like MD simulations (say $N \sim 10^5-10^6$ frames),
the $\mathcal{O}(N^3)$ complexity of the GP prior becomes prohibitively 
expensive, because the underlying GP-regression requires the inversion of the
full kernel matrix capturing temporal correlations across all $N$
data points, resulting in the cubic scaling.
Additionally, without further modifications, GP-VAEs cannot leverage 
batch training, as temporal correlation requires processing the 
entire trajectory simultaneously through the kernel matrix.
Fortunately, many works have provided the necessary tools to deal with 
these problems. 
Following Refs.~\citenum{Titsias09,Hensman13,Casale18,Pearce20,Jazbec21},
we now summarize the derivation of a GP-VAE loss function, that 
1.) scales to $10^5-10^6$ data points and 2.) allows mini-batching.
Compared to a standard VAE, the reconstruction part 
$\mathbb{E}_{\bm{\tilde{z}}\sim \tilde{q}_{\phi}(\bm{\tilde{z}}|\bm{x})}\left[\log p_{\theta}(\bm{x}|\bm{\tilde{z}})\right]$
remains identical, which is why we only consider the KL
divergence term in Eq.~\eqref{eq:starting_point}
\begin{align*}
    \KL{q_{\phi}(\bm{z} | \bm{x}, t)}{p(\bm{z}|t)} = - \mathbb{E}_{q(\bm{z} | \bm{x}, t)}\bigg[ \ln \frac{p(\bm{z}|t)}{q_{\phi}(\bm{z} | \bm{x}, t)} \bigg],
\end{align*}
where both the approximate posterior $q_{\phi}(\bm{z}|\bm{x},t)$ and the 
prior $p(\bm{z}|t) = \mathcal{GP}\bigl[0, k_{\nu,\ell}(t,t^\prime)\bigr]$
are time-dependent in contrast to the classical VAE.
To make the inference tractable, we follow the variational approximation
introduced by Pearce,\cite{Pearce20} which factorizes the posterior into
\begin{align*}
    q_{\phi}(\bm{z}|\bm{x},t) = \frac{p(\bm{z}|t)\tilde{q}_{\phi}(\tilde{\bm{z}}|\bm{x})}{Z(\bm{x},t)}.
\end{align*}
This factorization separates the temporal GP prior $p(\bm{z}|t)$ from the 
data-driven part $\tilde{q}_{\phi}(\tilde{\bm{z}}|\bm{x})$.
The normalization constant $Z(\bm{x}, t) = \int d\tilde{\bm{z}} \,
p(\bm{z}|t) \tilde{q}(\tilde{\bm{z}}|\bm{x})$ ensures that
$q_{\phi}(\bm{z} | \bm{x}, t)$ is a valid probability distribution,
where $\bm{z}$ and $\bm{\tilde{z}}$ refer to the same latent 
variable, where the tilde distinguishes the encoder output from 
its GP-regularized counterpart.
Substituting this factorization into the KL divergence term yields
\begin{align*}
    \KL{q_{\phi}(\bm{z} | \bm{x}, t)}{p_{\theta}(\bm{z}|t)} = &\mathbb{E}_{q_{\phi}(\bm{z} | \bm{x}, t)}\big[ \ln \tilde{q}_{\phi}(\tilde{\bm{z}} | \bm{x}) \big]\\
    &- \ln Z(\bm{x}, t),
\end{align*}
which allows us to rewrite Eq.~\eqref{eq:starting_point} as 
\begin{align}
    \mathcal{L} = &\mathbb{E}_{\tilde{q}_{\phi}(\bm{z} | \bm{x})} \big[ \ln p_{\theta}(\bm{x} | \bm{\tilde{z}}) \big] \nonumber \\
    &- \beta \big[ \mathbb{E}_{q_{\phi}(\bm{z} | \bm{x}, t)} \big[ \ln \tilde{q}_{\phi}(\tilde{\bm{z}} | \bm{x}) \big]
    - \ln Z(\bm{x}, t) \big].
\label{eq:gp_vae_loss}
\end{align}
The direct computation of Eq.~\eqref{eq:gp_vae_loss} is still computationally
prohibitive, since both the expectation $\mathbb{E}_{q_{\phi}(\bm{z} | \bm{x},
t)}\big[ \ln \tilde{q}_{\phi}(\tilde{\bm{z}} | \bm{x}) \big]$ as well as the
normalization constant $Z(\bm{x}, t)$ involve operations on the full kernel
matrix that scale as $\mathcal{O}(N^3)$. Therefore, it is necessary to use
sparse approximations for the GP,\cite{Quinonero05,Titsias09,Hensman13} which employ a
reduced set of $n_u \ll N$ inducing points with vectors $\bm{U} =
[\bm{u}_1, \dots, \bm{u}_{n_u}] \in \mathbb{R}^{n_u\times d_u}$ that are 
representative of the data. 
The dimensionality of an individual inducing point is denoted by $d_u$; 
in our setup, $d_u = 1$ since the inducing points correspond to time points.
The premise here is that a GP regression based on the inducing points
faithfully approximates the full GP regression over all $N$ data points.
Details of the derivation of this sparse approximation approach 
are given in the Appendix. 

Ultimately, this framework approximates the posterior mean $\bm{m}$ and 
covariance $\bm{B}$ for the GP latent representation at all
$N$ data points, resulting in the final (batch-wise) 
ELBO loss term\cite{Jazbec21}
\begin{align}
    \mathcal{L}_{\text{GP-VAE}} \equiv &\underbrace{\mathbb{E}_{\tilde{q}_{\phi}(\bm{\tilde{z}}|\bm{x})}\big[ \ln p_{\theta}(\bm{x}|\bm{\tilde{z}}) \big]}_{\text{reconstruction}} \label{eq:GP_elbo} \\
    &- \beta \underbrace{\left( \text{CE}\left[\mathcal{N}(\bm{m}, \bm{B}) \| \mathcal{N}(\boldsymbol{\tilde{\mu}}, \boldsymbol{\tilde{\sigma}}^2)\right] - \frac{n_b}{N} \mathcal{L}_{\text{H}} \right)}_{\text{GP regularization}}, \nonumber
\end{align}
which is the final result for the loss function of the GP-VAE model.
Here, $q_{\phi}(\bm{z}|\bm{x}, t) = \mathcal{N}(\bm{m}, \bm{B})$ 
and $n_b$ represent the GP-regularized posterior distribution and 
the batch size, respectively. $\mathcal{L}_{\text{H}}$ is a 
ELBO of the normalization term $\ln Z(\bm{x},t)$ in Eq.~\eqref{eq:gp_vae_loss}
introduced in Ref.~\citenum{Hensman13} that enables variational
sparse GP regression with mini-batching.

Compared to a vanilla VAE, the reconstruction part, i.e. the 
encoder and decoder parts of the  network in Eq.~\eqref{eq:GP_elbo}, 
remain unaltered.
However, the regularization term now compares the GP-based posterior 
$\mathcal{N}(\bm{m}, \bm{B})$ against the vanilla VAE posterior 
$\mathcal{N}(\bm{\tilde{\mu}}, \bm{\tilde{\sigma}}^2)$
by computing the dissimilarity through the cross-entropy (CE).
In other words, it penalizes deviations from the imposed correlation structure
implied by our choice of the GP kernel $k_{\nu}(t,t^\prime)$.
By using sparse GP regression techniques with inducing points, the 
computational complexity can be reduced from $\mathcal{O}(N^3)$ to 
$\mathcal{O}(n_b n_u^2+n_u^3)$.\cite{Jazbec21}
As a result, we obtain a time-aware
generative model to learn richer, physically well-motivated collective 
variables even for large data sets in the order of $\sim10^5-10^6$ points.

\section{Three-Dimensional Toy Model}
\begin{figure}[t]
    \centering
    \includegraphics[width=0.95\linewidth]{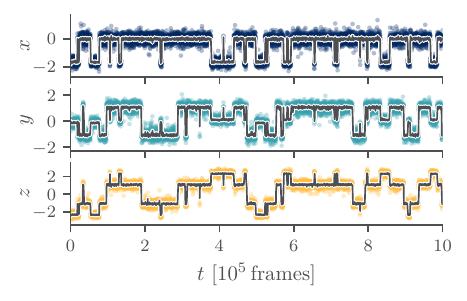}
    \caption{
        Time trace obtained from a Langevin simulation of the potential $\Phi(x,y,z)$
        described by Eq.~\eqref{eq:toypotential}. 
        The simulation was carried out for $10^6$ simulation steps using a 
        time step of $\Delta t = 5 \cdot 10^{-3}$, 
        a friction coefficient of $\gamma=1$ 
        and a temperature of $T=1$ (in dimension-less units).
    }
    \label{fig:toymodel_timetrace}
\end{figure}

To illustrate the effectiveness of our model and to test how well 
it can handle non-Markovian data, we simulated a three-dimensional trajectory $\bm{x}_{t}$
using the overdamped Langevin equation 
\begin{align*}
    \bm{x}_{t+1} = \bm{x}_t - \frac{\Delta t}{\gamma} \nabla \Phi(x,y,z) + \sqrt{\frac{2 k_\text{B} T \Delta t}{\gamma}}\xi_t,
\end{align*}
where $\gamma$ denotes the friction constant and $\xi_t$ represents Gaussian white noise drawn from a standard 
normal distribution with zero mean and unit variance.
The analytical toy potential used in our simulation
\begin{align}
    \Phi(x,y,z) = {} & -11.5 \big[e^{-x^2 -(y + 1.2)^2 - (z + 1.2)^2} \nonumber \\  
    & \quad + e^{-x^2 -(y - 1.2)^2 - (z - 1.2)^2} \big] \nonumber \\ 
    & -17 \big[ e^{-(x + 1.8)^2 - (y + 0.12)^2 - (z + 2.5)^2} \nonumber \\
    & \quad + e^{-(x + 1.8)^2 - (y - 0.12)^2 - (z - 2.5)^2}\big] \nonumber \\
    & + x^2 + y^2 + z^2
\label{eq:toypotential}
\end{align}
is designed to feature four well-separated basins, denoted as states 1 - 4.
Simulating a trajectory for $10^6$ steps,
Fig.~\ref{fig:toymodel_timetrace} displays the three-dimensional time trace.

\begin{figure*}
    \centering
    \includegraphics[width=0.85\linewidth]{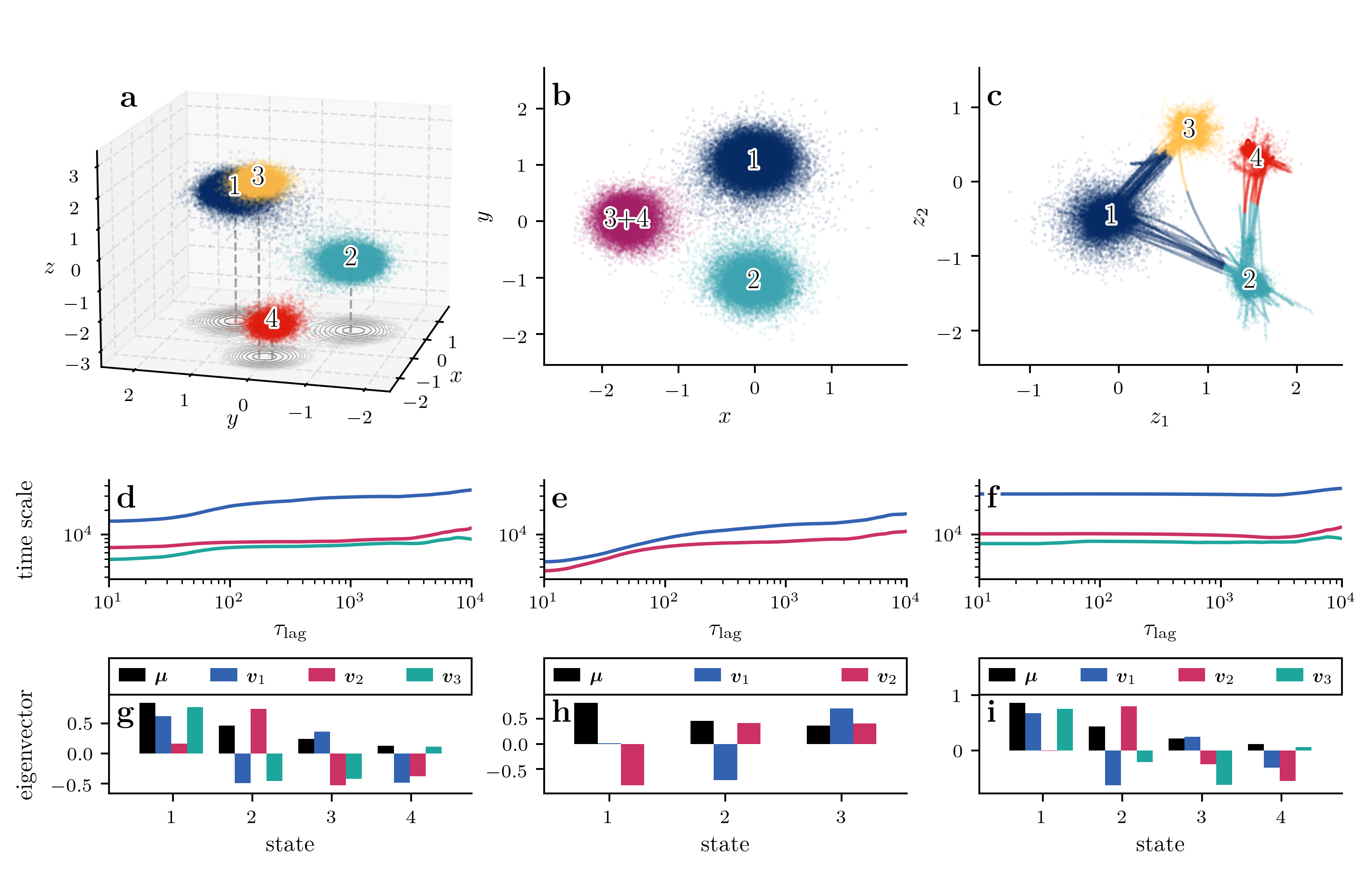}
    \caption{
        (a) Three-dimensional representation of the toy potential showing four
        distinct basins, labeled as state 1 (blue), 2 (cyan), 3 (yellow), and 4
        (red), respectively. 
        The contour lines in the $xy-$plane indicate the potential depth.
        (b) When projected onto the $xy$-plane, states 3 and 4
        overlap, making them indistinguishable
        without additional information. 
        (c) Using only the $xy$-plane data combined with time information, 
        the GP-VAE is capable of distinguishing the original state 3 and 4
        in the latent embedding $(z_1,z_2)$, where $z_1$ and $z_2$ denote
        latent coordinates (distinct from the physical Euclidean 
        $z$-coordinate), effectively recovering the Markovian dynamics.
        Lower panels (d-i) show column-wise MSM analyis results for each 
        scenario above: (d,g) correspond to the 3D case (a), (e,h) to the 
        $xy$-projection (b), and (f,i) to the GP-VAE embedding in (c). 
        Shown are the corresponding implied timescales, as well as 
        the eigenvectors $\bm{v}$ and the stationary distribution $\bm{\mu}$.
    }
    \label{fig:toymodel}
\end{figure*}

According to their definition in Fig.~\ref{fig:toymodel}a, states 2 and 3 are distinctly separated in all three dimensions, while
states 3 and 4 are only distinguishable along the $z$-axis, with
almost identical $x-$ and $y-$ values (see projection in the $xy-$plane in Fig.~\ref{fig:toymodel}b).
Dynamically, however, the states 3 and 4 are well separated with no direct 
transitions between them as transition must occur via the states 1 and 2.

In three dimensions, the system exhibits Markovian dynamics
since we used a Markovian Langevin equation to simulate the trajectory.
However, when we project the dynamics in the $xy-$plane, the states
3 and 4 overlap, which results in a loss of Markovianity in this
reduced space (see state 3+4 in Fig.~\ref{fig:toymodel}b).
The loss of Markovianity is reflected in the fact that the transition 
probabilities $p(i|j=$3+4) now depend on the history, i.e. whether 
state 3+4 is entered via state 1 or 2. 
This renders employing a traditional Markov state model (MSM) inappropriate.

\begin{figure}[b]
    \centering
    \includegraphics[width=0.99\linewidth]{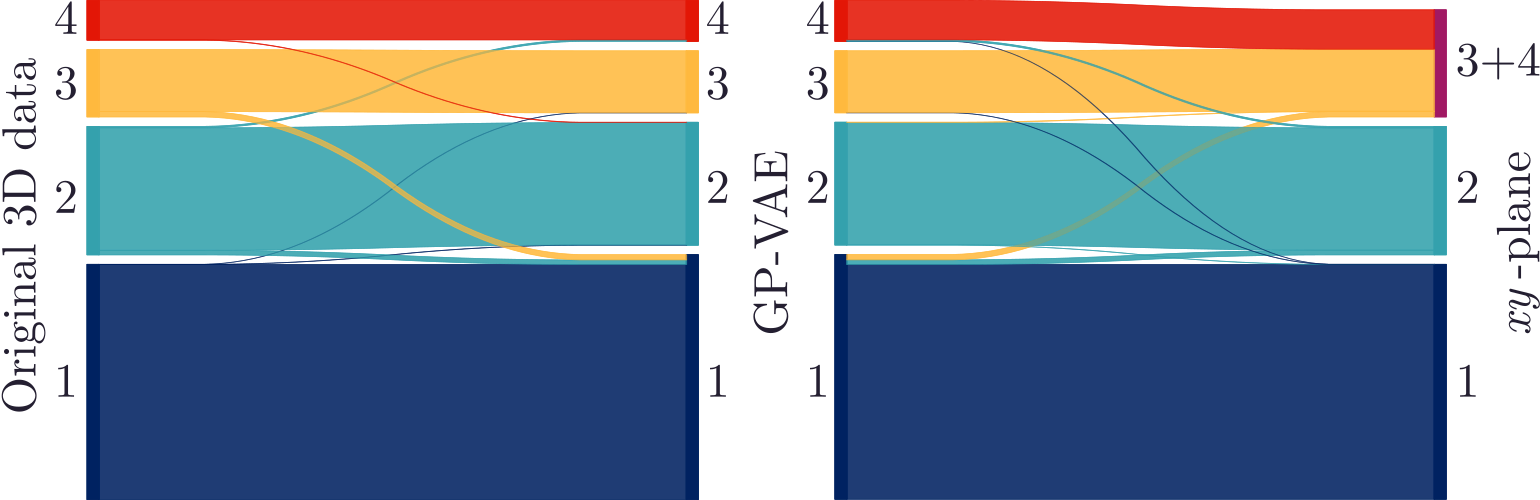}
    \caption{
        Sankey diagram illustrating the overlap between the states
        in the original 3D data (left), those obtained by clustering 
        the latent embedding of the GP-VAE (center) and the $xy-$plane (right).
        The width of each band indicate the fraction of frames in which both
        corresponding states temporally coincide.
        In the GP-VAE embedding, all four states are largely preserved
        with only minor differences in the transition region of state 1 
        (which might stem from dynamical coring).
    }
    \label{fig:overlap}
\end{figure}

To address this challenge and to enable the construction of a MSM,
our GP-VAE leverages both spatial data exclusively from the $xy-$plane 
as well as temporal information from the simulated trajectory to recover
the underlying Markovian dynamics.
Using $n_u=89$ inducing points identified by a change point detection 
algorithm also solely using data from the $xy-$plane 
(for more details see SM, Sec.~\SMsecInducing \ and 
Fig.~\SMinducing),
we trained our model using the parameters specified in supplementary 
material, Sec.~\SMsecParameters.  
For both the encoder and decoder, we use five
hidden layers with widths of 10, 32, 64, 32, 10 units, respectively. 
To determine an appropriate kernel length scale parameter 
$\ell$ for the Matérn kernel, we estimate the characteristic lifetime of 
metastable states from the $x$-coordinate trajectory.
Identifying $\sim 14$ distinct stable sequences while filtering out fast
fluctuations, we calculate the average state lifetime as the total
trajectory length divided by the number of transitions 
$\ell = 10^6/13 \approx 7.5\cdot 10^4$.
This parameter choice ensures strong temporal correlations within stable 
sequences while minimizing correlations between distinct metastable states.
Fig.~\ref{fig:toymodel}c shows that the model
indeed is able to separate the overlapping state 3+4 into two distinct states,
which show excellent agreement with the original states 3 and 4
in the full three-dimensional space.
For a better visualization, Fig.~\ref{fig:overlap} shows a Sankey diagram illustrating the
temporal overlap between the states in the full and reduced space with 
the GP-VAE embedding.
Only minor deviations are notable for state 1 which are arguably resulting
from the dynamical coring that we carried out before constructing a 
MSM for further analysis.

This successful separation of geometrically indistinguishable
states is indeed attributable to the temporal correlations captured by
the Matérn kernel, as demonstrated by a comparison with a vanilla VAE
that solely relies on geometric information. Since both overlapping 
states 3 and 4 are geometrically not distinguishable, a standard 
VAE approach fails to distinguish them, as the resulting embedding
in Fig.~\SMvae~proves.

To construct the MSMs, we used $k$-means clustering ($k = 1000$), 
followed by MPP lumping and (iterative) dynamical coring with a lag time 
($\tau_{\text{lag}} = \tau_{\text{cor}} = 10$ frames).\cite{Jain14, Nagel19}
After computing the transition matrices $\bm{T}(\tau)$ 
for all three MSMs, we calculated the implied time scales 
$t_i(\tau) = -\tau/\log \lambda_i(\tau)$,
where $\lambda_i$ is the eigenvalue corresponding to the eigenvector
$\bm{v}_i$ of $\bm{T}(\tau)$.
All MSM related analyses were performed using msmhelper.\cite{Nagel23a}
The results are shown in Figs.~\ref{fig:toymodel}d-f and g-i, respectively. 
Interestingly, the implied timescales of both the full three-dimensional
model and the GP-VAE embedding converge towards identical
values for large $\tau_\text{lag}$. 
Notably, the GP-VAE embedding achieves much faster convergence, which
we attribute to the inherent Markovianity imposed by the usage
of Markovian kernels such as e.g. the employed Matérn-kernel. 
The GP regression in the latent space effectively acts as a 
time-aware filtering that filters out non-Markovian noise components 
and hence results in microstates with significantly higher 
metastabilities. This is evident in the dendrograms 
illustrating the hierarchical lumping towards macrostates 
(see Fig.~\SMdendrograms).

To further assess the quality of the latent embedding of our model,
we compare the eigenvectors of the two MSMs in the full space and in
the reproduced GP space. This comparison reveals that the GP-VAE
successfully captures the system's full dynamical structure even though
it has only limited information through the $xy-$plane.
Consequently, the equilibrium state population (i.e. the stationary
probability distribution $\boldsymbol{\mu}$) from the GP-VAE matches that of the 
full three-dimensional model.
Also, the first eigenvector $\bm{v}_1$, that characterizes the slowest 
dynamical process in the system, is faithfully reproduced, which  
highlights the GP-VAE's ability to capture dominant transition pathways
between metastable conformations.
Higher-order eigenvectors $\bm{v}_2$ and $\bm{v}_3$, which represent faster
dynamical processes, remain in qualitative agreement with the original
model; however, with some expected deviations due to the lack of knowledge
of the original dynamics in the $z-$dimension.

In summary, the model demonstrated the remarkable 
ability to extract meaningful collective variables even when the 
input data might miss essential geometric information due to 
erroneous steps in previous dimensionality reduction.
The GP-VAE not only recovered the correct state assignment from
incomplete spatial information but also achieved a faster convergence
of the implied timescales in subsequent MSM construction. 
We attribute this to two key factors: first, the inherent 
mathematical properties of a Markovian kernel
impose a Markovian structure on the latent space by design.
Secondly, the GP regression---particularly when smooth
kernels such as the Matérn kernel are used---acts as a time-aware
low-pass filter.\cite{Rasmussen06}
Consequently, slowly varying and kinetically relevant modes are 
favored while high-frequency, non-Markovian noise components 
are filtered out.
Both factors combined might not only facilitate the subsequent 
construction of MSMs, but can also help to disentangle distinct
dynamical processes, even when they are geometrically
indistinguishable.

\section{Functional Dynamics of T4 Lysozyme}

\begin{figure}[b]
    \centering
    \includegraphics[width=0.95\linewidth]{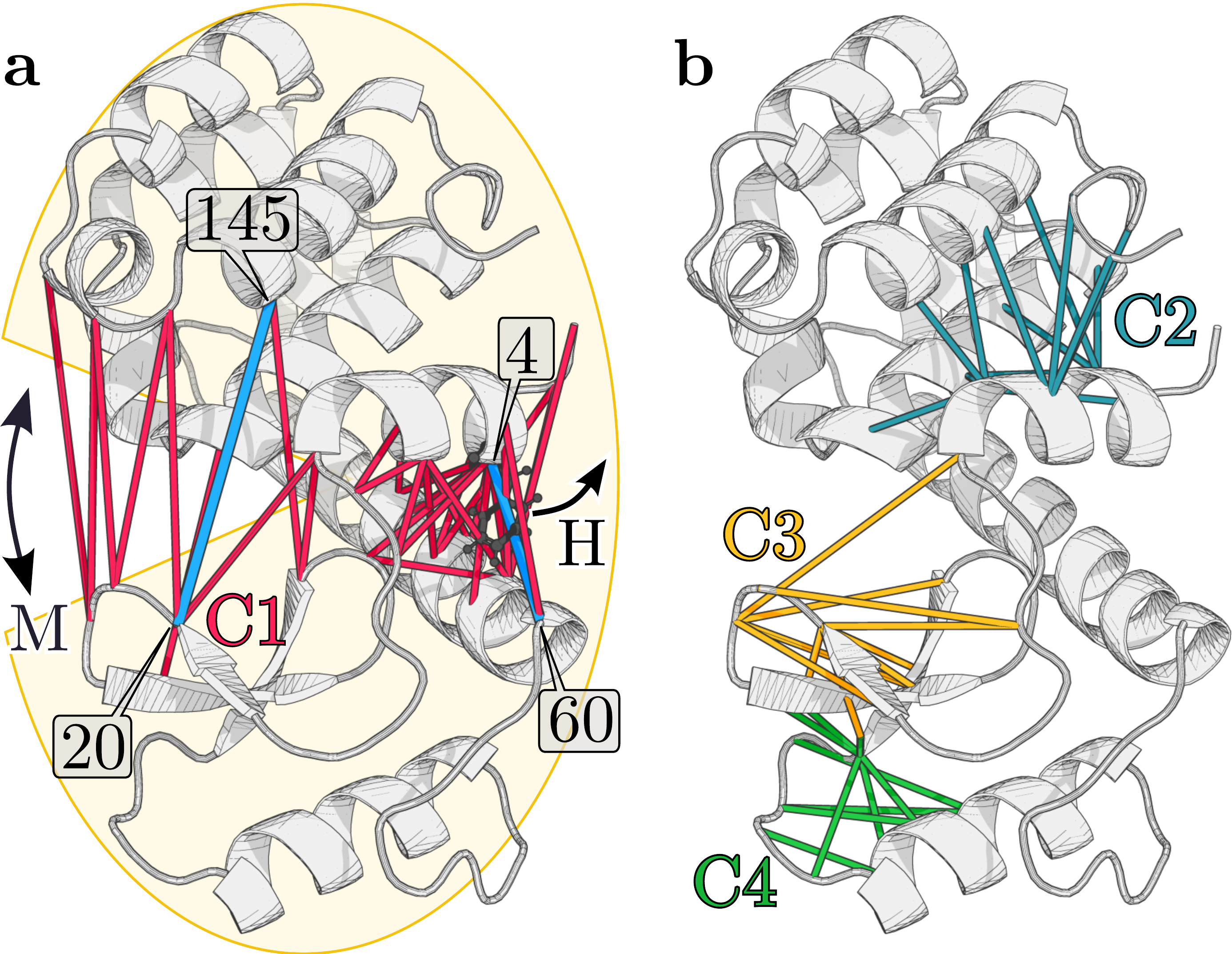}
    \caption{
    Structure of T4L, indicating MoSAIC clusters of inter-residue contacts shown as red (C1), blue (C2), yellow (C3) and green (C4) lines. 
    (a) Cluster C1 accounts for the open$\leftrightarrow$closed motion of T4L, spanning from the hinge region (H) to the mouth (M) region,
    with the most important coordinates, $d_{4,60}$ 
        and $d_{22,137}$ highlighted in blue. 
        (b) The remaining three clusters, C2-C4, describe different processes
        that are not directly linked to the open$\leftrightarrow$closed
        motion.
    }
    \label{fig:t4l_mosaic}
\end{figure}

Following this proof of principle, we wish to apply the GP-VAE method to 
all-atom MD data of functional protein dynamics.
To this end, we consider a previously performed $50\,\mu$s-long MD simulation
of T4 lysozyme,\cite{Ernst17} which was already analyzed in some detail in Refs.\ \onlinecite{Post22a, Nagel24}. T4L is a 164-residue enzyme that performs an 
open$\leftrightarrow$closed transition of its two domains, which
resembles the motion of a Pac-Man, see Fig.~\ref{fig:t4l_mosaic}a. Within 
the simulation time of $50\,\mu$s, about ten open$\leftrightarrow$closed transitions were observed, see Fig.~\SMlysozymeinducing. 
In the following, we show that a GP-VAE analysis on T4L reveals dynamically distinct states (that appear identical in a conventional analysis), and identify 
conformational prerequisites for the functional motion that could not
detected before.\cite{Post22a, Nagel24}

\subsection{MoSAIC Analysis}
To systematically analyze T4L's functional dynamics, we 
focus on inter-residue contact distances. We assume that a contact is
formed if the distance $r_{ij}$ between the closest non-hydrogen atoms
of residues $i$ and $j$ is shorter than 4.5\,\AA, and if the 
contact is formed more than 1\,\% of the simulation time.\cite{Ernst15}
To discriminate collective motions underlying functional dynamics from
uncorrelated motion, we calculated the
correlation matrix of these coordinates and rearranged this matrix in
an approximately block-diagonal form, employing the Python package
MoSAIC.\cite{Diez22} 
As shown in the SM (Fig.~\SMlysozymemosaicmatrix~and Table~\SMtabmosaic), 
this yields four clusters (C1 to C4) of highly correlated contacts, 
where the largest cluster C1 describes the open$\leftrightarrow$closed
transition. As shown in Fig.~\ref{fig:t4l_mosaic}a, the corresponding distances span from the hinge region to the mouth
region of T4L. While the exact molecular mechanism is quite 
intricate,\cite{Post22a} the open$\leftrightarrow$closed transition 
can be characterized by a simple model using two key inter-residue distances:
$d_{20,145}$ (between Asp20 and Arg145) monitoring the opening of the mouth, while $d_{4,60}$ (between Phe4 and Lys60) describing the dynamics of the hinge 
region. That is, the side chain of Phe4 transitions 
between solvent-exposed and hydrophobically-buried states, 
effectively acting as a locking mechanism.\cite{Ernst17,Post22a}
This results in a free energy landscape $\Delta G(d_{4,60},d_{20,145})$ with two distinct minima corresponding to the open and closed conformations
(Fig.~\ref{fig:t4l_embeddings}a).

The remaining three smaller MoSAIC clusters, C2-C4, describe distinct 
conformational processes in different regions of T4L that are not 
directly correlated to the primary open$\leftrightarrow$closed transition 
(see Fig.~\ref{fig:t4l_mosaic}b and Fig.~\SMlysozymemosaicmatrix). Using GP-VAE, we nevertheless find a hitherto unknown dynamical coupling between C1 and C3. 

\subsection{GP-VAE Embedding of the T4L Reaction Coordinates}

\begin{figure}[t]
    \centering
    \includegraphics[width=1\linewidth]{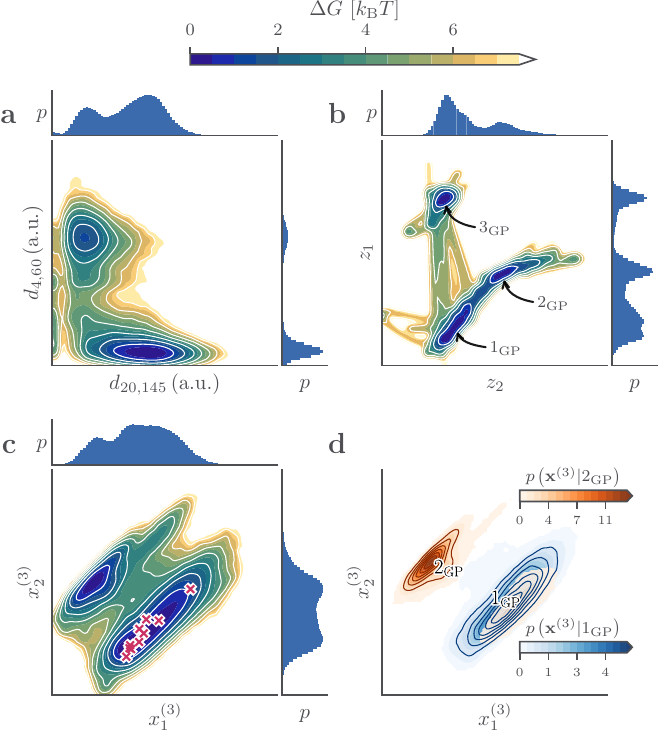}
    \caption{   
        Two-dimensional free energy landscapes of T4L, with marginal probability distributions shown along each axis. (a) Energy landscape 
        constructed from the two key distances, $d_{4,60}$ and $d_{20,145}$,
        showing a clear two-state behavior.
        (b) The landscape of the corresponding GP-VAE embedding reveals a second "open" state ($2_\text{GP}$) that is dynamically disconnected from the open$\leftrightarrow$closed region. 
        (c) Energy landscape of MoSAIC cluster C3, obtained from a principal component analyses of the coordinates of this cluster. The first two
        components $(x_1^{(3)}$ and $x_2^{(3)})$ are found to clearly separate 
        the two GP states $1_\text{GP}$ and $2_\text{GP}$. Red crosses indicate
        the points of the the open$\leftrightarrow$closed transitions.
        (d) Probability distributions of states $1_\text{GP}$ and $2_\text{GP}$ projected onto $(x_1^{(3)}, x_2^{(3)})$ space.
    }
    \label{fig:t4l_embeddings}
\end{figure}

To investigate whether there are hidden degrees of freedom, 
we apply the GP-VAE to the reaction coordinates $d_{4,60}$ and $d_{20,145}$, using the hyperparameters specified in Tab.~\SMtabhyperparameters.
Given that the average lifetime of the open and closed states is $\sim 5\,\mu$s, we set the length scale $\ell$ of the Matérn kernel moderately longer to $7\,\mu$s. (In fact, we found that kernel length scales in the range of 
$\ell \sim (2-10)\,\mu$s generally work well, indicating that the method 
is robust in this regard.)
The resulting embedding shown in Fig.~\ref{fig:t4l_embeddings}b
reveals three distinct states: the previously single open state
now separates into two open "GP" states, $1_\text{GP}$ and $2_\text{GP}$,
while $3_\text{GP}$ corresponds to the closed conformation.
By imposing temporal correlations through the Matérn kernel,
a part of the open state in $\Delta G(d_{4,60},d_{20,145})$
splits off, as it is temporally completely uncorrelated to the open
conformation of T4L. 
Consequently, conformational transitions between the open and closed 
T4L conformations proceed exclusively through state $1_\text{GP}$, 
whereas state $2_\text{GP}$ is kinetically isolated.
This newly discovered $2_\text{GP}$ state---emerging only through
incorporation of temporal information and remaining undetectable in
purely static structural analysis---requires validation of its
physical significance.

To explore a possible connection of the open$\leftrightarrow$closed dynamics of cluster C1 with the seemingly uncorrelated cluster C2, C3 and C4, we performed 
principal component analyses of each of these three clusters. In the case of C3, we indeed found that the free energy landscape exhibits two states, reflecting the two GP states $1_\text{GP}$ and $2_\text{GP}$, see Fig.~\ref{fig:t4l_embeddings}c. 
Hence, the GP-VAE embedding reveals an unexpected functional relationship,
although the correlation matrix (Fig.~\SMlysozymemosaicmatrix)
shows no significant cross-correlations between clusters C1 and C3.
To test this hypothesis, we project the probability distributions 
of the states $1_\text{GP}$ and $2_\text{GP}$ into the 
$(x_1^{(3)}, x_2^{(3)})$ space (see Fig.~\ref{fig:t4l_embeddings}d). 
As anticipated, both states clearly map to one of the two main basins within this principal component subspace.
To further validate this connection, we mark the points of 
all open$\leftrightarrow$closed transition 
by red crosses in Fig.~\ref{fig:t4l_embeddings}c. This confirms that
transitions occur exclusively when cluster C3 occupies the region corresponding to 
state $1_\text{GP}$, while the second state $2_\text{GP}$ represents a 
dynamically isolated open sub-state.

A structural analysis of both GP states is provided in the SM, where Fig.~\SMlysozymejawstructure~displays the 
lowest free energy structures of each state and 
Fig.~\SMlysozymehistograms~shows the probability distributions of 
the main inter-residue distances that distinguish the two states.
Ultimately, only state $1_\text{GP}$ facilitates the highly
coordinated and simultaneous switching of coordinates contained 
in C1, which is necessary for the open$\leftrightarrow$closed
transition of T4L.

Finally, we want to stress that our analysis so far focused on two-dimensional
input data, that is, distances $d_{4,60}$ and $d_{20,145}$. To show that
the GP-VAE can also handle high-dimensional input data,
we embedded all 32 contact distances from 
MoSAIC cluster 1 (Fig.~\SMlysozymeembedding), yielding a latent 
space that preserves the key conformational features identified 
in the two-dimensional case. 
This confirms the dimensional robustness of our approach and 
indicates its potential as a powerful tool for dimensionality 
reduction.

\section{Conclusion}
\vspace{-4mm}

We have introduced a physics-informed representation learning 
framework for MD simulations, which naturally 
accounts for the temporal structure of trajectory data.
In contrast to traditional methods which operate under the i.i.d.\ assumption,
the proposed Gaussian Process Variational Autoencoder (GP-VAE) framework
explicitly models time correlations in the latent space via
specialized kernel functions.
The approach appears to be especially helpful in cases where essential degrees 
of freedom remain unobserved in the initial feature selection.

Central to our method is the use of Markovian kernel functions, 
such as e.g. the Matérn kernel, which allows preserving the Markovianity
of the input features during dimensionality reduction.
This kernel choice explicitly enforces temporal correlations
in the latent space by acting as a time-aware smoothing of the data, 
which filters out non-Markovian noise components.
By explicitly designing the feature extraction process to enforce
Markovian properties in the latent space---through physics-informed
kernel choice and time-aware smoothing---the resulting embeddings
represent an optimal starting point for the construction of 
Markov state models:
the embedding exhibits significantly enhanced metastability
of the microstates allowing for smaller lag times and enabling 
faster convergence of implied timescales.
The positive effects of a temporal filtering were already indicated
in previous works, where a low-pass filtering
of the MD data was employed, suppressing fast non-Markovian fluctuation
and thereby reducing the misclassification of frames in the 
transition regions.\cite{Nagel23b}

Employing a four-state toy model as a proof of principle, we have demonstrated that time-aware embeddings can separate dynamically distinct states that appear
geometrically similar, due to the lack of deciding distinctive feature.
The GP-VAE successfully recovered the 
correct state assignments by leveraging temporal correlations 
and thus was able to correctly separate and distinguish the two 
overlapping states without having observed their separation in space.  
Moreover, the model is able
to precisely recover the dynamics from the original three-dimensional 
space,
as demonstrated by the MSM.

Applying the GP-VAE to T4 lysozyme,  we identified a previously 
unrecognized dynamically isolated open substate, that is functionally 
coupled with MoSAIC cluster 3.
The open$\leftrightarrow$closed transition occurs conditionally on the 
protein being in one of the two metastable states of this cluster,
which explains that this coupling cannot be detected by 
instantaneous correlations alone.
This finding highlights the potential of our approach to uncover
temporal and possibly causal links between different functional
subunits of proteins, as e.g. revealed by MoSAIC.

Although we deliberately selected two-dimensional examples to clearly 
illustrate the role of the temporally-aware GP prior, it is 
important to stress that this framework can embed data of arbitrary 
dimensionality into latent spaces of any desired dimension.

\subsection*{Supplementary Material}
\vspace{-4mm}
Includes additional information on the
Matérn kernel, the change point detection algorithm for selecting
inducing points, and model hyperparameters used for GP-VAE.
For the toy model, it contains details about the dynamical
coarse graining of microstates, the chosen inducing points,
and compares the GP-VAE embedding to a vanilla VAE.
For T4 lysozyme, it includes the results of the MoSAIC feature selection,
the inducing points choices, and a structural analysis of the 
newly identified substates.
\vspace{-4mm}

\subsection*{Software}
\vspace{-4mm}
The code for the GP-VAE is implemented using PyTorch\cite{Paszke19}
and is freely available for public use at \url{https://github.com/moldyn/GP-TEMPEST}.
\vspace{-4mm}

\subsection*{Acknowledgment}
\vspace{-4mm}
The authors thank Marius Lange for helpful comments and discussions.
This work has been supported by the Deutsche
Forschungsgemeinschaft (DFG) within the framework of the Research Unit FOR 5099
“Reducing complexity of nonequilibrium”(project No. 431945604), the High
Performance and Cloud Computing Group at the Zentrum für Datenverarbeitung of
the University of Tübingen and the Rechenzentrum of the University of Freiburg,
the state of Baden-Württemberg through bwHPC and the DFG through grant no INST
37/935-1 FUGG (RV bw16I016), and the Black Forest Grid Initiative.
\vspace{-4mm}

\section*{Appendix}
\vspace{-4mm}
Starting with Eq.~\eqref{eq:gp_vae_loss}, here we provide some details
of the derivation of Eq.~\eqref{eq:GP_elbo} using sparse approximations
for the GP regression.\cite{Quinonero05,Titsias09,Hensman13}

To this end, we introduce a reduced set of $n_u \ll N$ inducing points with vectors
$\bm{U} = [\bm{u}_1, \dots, \bm{u}_{n_u}] \in \mathbb{R}^{n_u\times d_u}$---where 
in our setting $d_u=1$ as the inducing points correspond to time---enables
GP regression based on this subset of representative points.
We denote the GP evaluated at the inducing points by 
$\bm{f}_u~\equiv~f(\bm{U})~\sim \mathcal{N}(f(\bm{U})|\boldsymbol{\mu}, \bm{A})$,
which approximates the GP regression on all data points, $\bm{f}_N$.
Just like in standard GP regression, $\bm{f}_u$ is modeled as 
a multivariate Gaussian with mean vector $\bm{\mu} \in \mathbb{R}^{n_u}$ 
and covariance matrix $\bm{A} \in \mathbb{R}^{n_u \times n_u}$.
For a introduction to GPs, we recommend Ref.~\citenum{Rasmussen03}.
The distribution $p(\bm{f}_u) = \mathcal{N}(\bm{f}_u | \bm{0}, \bm{K}_{uu})$
serves as the GP prior over the inducing points, while a variational posterior
$q_{\phi}(\bm{f}_u | \bm{x}, t) = \mathcal{N}(\bm{\mu}, \bm{A})$ uses freely optimized
mean $\bm{\mu}$ and covariance $\bm{A}$ 
since both depend on the output of the vanilla VAE posterior 
$\tilde{q}_{\phi}(\bm{\tilde{z}}|\bm{x})$. 

Specifically, based on Titsias work,\cite{Titsias09} Jazbec and coworkers 
suggested computing these (intermediate) variational quantities 
at the inducing points $u$ as stochastic estimates for each 
latent dimension $l \in \{1, \dots, L\}$ and for each batch 
$b$\cite{Jazbec21}
\begin{align*}
    \bm{\mu}^l_{b} &= \frac{N}{n_b} \bm{K}_{uu} \left(\bm{\Sigma}^l_{b}\right)^{-1} \bm{K}_{ub} \text{diag}\left(\tilde{\bm{\sigma}}_{b}^{-2}\right) \tilde{\bm{\mu}}^l_{b},\\
    \bm{A}^l_{b} &= \bm{K}_{uu} \left(\bm{\Sigma}^l_{b}\right)^{-1} \bm{K}_{uu},\\
    \intertext{where $\bm{\Sigma}^l_{b}$ is given as}
    \bm{\Sigma}^l_{b} &= \bm{K}_{uu} + \frac{N}{n_b} \bm{K}_{ub} \text{diag}\left(\tilde{\bm{\sigma}}^{-2}_{b}\right) \bm{K}_{bu}.
\end{align*}
In this expression, $\bm{K}_{ub} = k_{\nu,l}(\bm{U}, \bm{t}_b) \in \mathbb{R}^{n_u \times n_b}$ represents the kernel matrix computed between the $n_u$ inducing
points $\bm{U}$ and the $n_b$ data points $\bm{t}_b$ within the
current batch $b$.
Importantly, these estimators converge to the true values for very large
batch sizes $n_b\rightarrow N$.
Following Refs.~\citenum{Titsias09,Hensman13}, we obtain the final posterior
distribution parameters in the form of the posterior mean $\bm{m}$ and
covariance $\bm{B}$ of the GP latent embedding at all data
points in the batch $b$:
\begin{align*}
    \bm{m}^l_{b} &= \frac{N}{n_b} \bm{K}_{bu} \left(\bm{\Sigma}^l_{b}\right)^{-1} \bm{K}_{ub} \text{diag}\left(\tilde{\bm{\sigma}}^{-2}_{b}\right) \tilde{\bm{\mu}}^l_{b},\\
    \bm{B}^l_b &= \text{diag} \left(\bm{K}_{bb} - \bm{K}_{bu} \bm{K}_{uu}^{-1} \bm{K}_{ub} + \bm{K}_{bu} \left(\bm{\Sigma}^{l}_{b}\right)^{-1} \bm{K}_{ub}\right).
\end{align*}
Since we assumed independence across latent dimensions, we can
characterize the posterior distribution of the latent embedding
\begin{align*}
    q_{\phi}(\bm{z} | \bm{x}, t) = \prod_{l=1}^L q_{\phi}(\bm{z}^l | \bm{x}, t) = \mathcal{N} (\bm{m}, \bm{B}).
\end{align*}
This allows us to calculate the term 
$\mathbb{E}_{q_{\phi}(\bm{z} | \bm{x}, t)}\big[ \ln \tilde{q}_{\phi}(\tilde{\bm{z}} | \bm{x})\big]$
in Eq.~\eqref{eq:gp_vae_loss} as
\begin{align*}
  \mathbb{E}_{q_{\phi}(\bm{z}|\bm{x}, t)}\big[ \ln \tilde{q}_{\phi}(\tilde{\bm{z}} | \bm{x}) \big]
  &= \int \mathrm{d}\bm{z}\, \mathcal{N}(\bm{z}|\bm{m}, \bm{B}) \ln \mathcal{N}(\bm{z} |\tilde{\bm{\mu}}, \tilde{\bm{\sigma}}^2) \\
  &= \mathrm{CE} \left[ \mathcal{N}(\bm{m}, \bm{B}) \,\Vert\, \mathcal{N}(\tilde{\bm{\mu}}, \tilde{\bm{\sigma}}^2) \right].
\end{align*}
The cross-entropy (CE) between the two Gaussian distributions 
$\mathcal{N}(\bm{m}, \bm{B})$ and $\mathcal{N}(\tilde{\bm{\mu}}, \tilde{\bm{\sigma}}^2)$
can be analytically computed as\cite{Tian24}
\begin{align}
    \text{CE} \left[ \cdot \Vert \cdot \right] &= \frac{1}{2} \left\{ L \ln 2 \pi + \ln \det \text{diag} (\tilde{\bm{\sigma}}^2) \right. \notag \\
    &\quad\left. +(\bm{m} - \tilde{\bm{\mu}})^\top \text{diag}(\tilde{\bm{\sigma}}^{-2}) (\bm{m} - \tilde{\bm{\mu}}) \right. \notag \\
    &\quad\left. + \text{tr} \left[ \text{diag}(\bm{B}) \text{diag}(\tilde{\bm{\sigma}}^{-2}) \right] \right\}, \label{SI:ML:eq:CE}
\end{align}
of which we can compute all quantities: $\tilde{\bm{\mu}}$
and $\tilde{\bm{\sigma}}^2$ are the variational parameters
learned by the classical VAE encoder and $\bm{m}$ and $\bm{B}$
are the posterior mean and covariance of the GP latent embedding
at all data points.

Lastly, the normalization term $Z(\bm{x}, t)$ remains to be calculated.
Recall from the factorization, that the direct calculation of
$Z(\bm{x}, t) = \int \diff \tilde{\bm{z}} \, p(\bm{z}|t) \tilde{q}_{\phi}(\tilde{\bm{z}} | \bm{x})$
is intractable, since it involves the full kernel matrix in $p(\bm{z}|t)$.
Following Hensman \textit{et al.} in Ref.~\citenum{Hensman13},
this intractability can be circumvented by constructing an 
ELBO that 1.) serves as a  tractable lower bound to $\ln Z(\bm{x}, t)$ and 
2.) can be computed using mini-batches
\begin{align*}
    \ln Z(\bm{x}, t) \geq \mathcal{L}_{\text{H}} = &\sum_{i=1}^N \left\{ \ln \mathcal{N} \left(\bm{\tilde{\mu}}_i | \bm{k}_i \bm{K}_{uu}^{-1} \bm{\mu}, \bm{\tilde{\sigma}}^2_i \right) \right. \\
    &\quad \left. - \frac{1}{2\bm{\tilde{\sigma}}^2_i} \left[ \tilde{k}_{ii} + \text{Tr}(\bm{A} \bm{\Lambda}_i) \right] \right\} \\
    & \quad - \KL{q_{\phi}(\bm{f}_u | \bm{x}, t)}{p(\bm{f}_u)}.
\end{align*}
Here, $\bm{k}_i$ represents the $i$-th row of $\bm{K}_{Nu}$,
$\tilde{k}_{ii}$ denotes the $i$-th diagonal element of
$\bm{K}_{NN} - \bm{K}_{Nu} \bm{K}_{uu}^{-1} \bm{K}_{uN}$,
This ELBO contains a KL divergence term between the variational
posterior and GP prior over the inducing variables.
To compute this tractably, we evaluate
\vspace{-0.5cm}
\begin{widetext}
\begin{align}
    \KL{q_{\phi}(\bm{f}_u|\bm{x}, t)}{p(\bm{f}_u)}
    &= \mathbb{E}_{q_{\phi}(\bm{f}_u|\bm{x}, t)} \left[ \ln \frac{q_{\phi}(\bm{f}_u|\bm{x}, t)}{p(\bm{f}_u)} \right]  \nonumber \\
    &= \mathbb{E}_{q_{\phi}(\bm{f}_u|\bm{x}, t)} \big[ \ln \mathcal{N}(\bm{f}_u | \boldsymbol{\mu}, \bm{A}) \big] - \mathbb{E}_{q_{\phi}(\bm{f}_u|\bm{x}, t)} \big[ \ln \mathcal{N}(\bm{f}_u| \bm{0}, \bm{K}_{uu}) \big]. \label{SI:eq:KL_div}
    \intertext{Using the standard expression for the log-probability of a Gaussian
distribution, the first term simplifies to}
\mathbb{E}_{q_{\phi}(\bm{f}_u|\bm{x}, t)} \left[ \ln \mathcal{N}(\bm{f}_u | \boldsymbol{\mu}, \bm{A}) \right]  \nonumber
    &= \mathbb{E}_{q_{\phi}(\bm{f}_u|\bm{x}, t)} \left[ -\frac{1}{2} \left( (\bm{f}_u - \boldsymbol{\mu})^\top \bm{A}^{-1} (\bm{f}_u - \boldsymbol{\mu}) + \ln \det \bm{A} + n_u \ln 2\pi \right) \right] \\
    &= -\frac{1}{2} \left( n_u + \ln \det \bm{A} + n_u \ln 2\pi \right). \label{SI:eq:KL_div_2}
    \intertext{Similarly, we can simplify the second term:}
    \mathbb{E}_{q_{\phi}(\bm{f}_u | \bm{x}, t)} \big[ \ln \mathcal{N}(\bm{f}_u | \bm{0}, \bm{K}_{uu}) \big] \nonumber
    &= \mathbb{E}_{q_{\phi}(\bm{f}_u | \bm{x}, t)} \left[ -\frac{1}{2} \left( \bm{f}_u^{\top} \bm{K}_{uu}^{-1} \bm{f}_u + \ln \det \bm{K}_{uu} + n_u \ln 2\pi \right) \right] \\
    &= -\frac{1}{2} \left[ \operatorname{tr} \left( \bm{K}_{uu}^{-1} \bm{A} \right) + \boldsymbol{\mu}^{\top} \bm{K}_{uu}^{-1} \boldsymbol{\mu} + \ln \det \bm{K}_{uu} + n_u \ln 2\pi \right].\label{SI:eq:KL_div_3}
    \intertext{where we used the expectation of the quadratic form of a Gaussian.
Substituting Eq.~\eqref{SI:eq:KL_div_2} and Eq.~\eqref{SI:eq:KL_div_3}
back into Eq.~\eqref{SI:eq:KL_div}, we obtain}
    \KL{q_{\phi}(\bm{f}_u|\bm{x}, t)}{p(\bm{f}_u)}
    &= \mathbb{E}_{q_{\phi}(\bm{f}_u | \bm{x}, t)} \big[ \ln \mathcal{N}(\bm{f}_u | \boldsymbol{\mu}, \bm{A}) \big] - \mathbb{E}_{q_{\phi}(\bm{f}_u | \bm{x}, t)} \big[ \ln \mathcal{N}(\bm{f}_u | \bm{0}, \bm{K}_{uu}) \big] \nonumber \\
    &= \frac{1}{2} \left[ -n_u + \operatorname{tr} \left( \bm{K}_{uu}^{-1} \bm{A} \right) + \boldsymbol{\mu}^{\top} \bm{K}_{uu}^{-1} \boldsymbol{\mu} + \ln \frac{\det \bm{K}_{uu}}{\det \bm{A}} \right]. \label{SI:eq:KL_div_final}
    \intertext{Having derived the KL divergence term analytically,
we can now return to the Hensman ELBO and reformulate it relying only
on computable quantities:} \nonumber
\end{align}
\vspace{-1.5cm}
\begin{align}
    \mathcal{L}_{\text{H}} &= \sum_{i=1}^N \left\{ \ln \mathcal{N} \left(\bm{\tilde{\mu}}_i | \bm{k}_i \bm{K}_{uu}^{-1} \bm{\mu}, \bm{\tilde{\sigma}}^2_i \right) - \frac{1}{2\tilde{\sigma}^2_i} \left[ \tilde{k}_{ii} + \text{Tr}(\bm{A} \bm{\Lambda}_i) \right] \right\} \notag \\
    &\quad - \frac{1}{2} \left[ - n_u + \text{tr}(\bm{K}_{uu}^{-1}\bm{A}) + \bm{\mu}^{\top}\bm{K}_{uu}^{-1}\bm{\mu} + \ln \frac{\det \bm{K}_{uu}}{\det \bm{A}} \right], \label{SI:ML:eq:L_H_complete}
\end{align}
where we have substituted our derived expression for the KL divergence from Eq.~\eqref{SI:eq:KL_div_final}.
This completes the derivation of all necessary components.
We can now return to Eq.~\eqref{eq:gp_vae_loss} and combine the 
following three tractable terms
\begin{itemize}
    \item[$\cdot$] reconstruction term (unchanged from standard VAE)
    \vspace{-2mm}
\item[$\cdot$] GP regularization: Cross-entropy term between sparse GP     \vspace{-2mm}
posterior and VAE encoder [Eq.~\eqref{SI:ML:eq:CE}]
    \item[$\cdot$] normalization: Hensman ELBO $\mathcal{L}_{\text{H}}$ [Eq.~\eqref{SI:ML:eq:L_H_complete}]
\end{itemize}
into the final loss function, serving as a lower bound
to the log evidence of the sparse GP-VAE model in 
Eq.~\eqref{eq:GP_elbo}.
\end{widetext}

\bibliographystyle{./Bib/STY/aip+title}
\bibliography{./Bib/stock.bib,./Bib/md.bib, new}

\begin{thebibliography}{10}

\bibitem{Berendsen07}
H.~J.~C. Berendsen,
\newblock {\em Simulating the Physical World},
\newblock Cambridge University Press, Cambridge, 2007.

\bibitem{Bolhuis00}
P.~G. Bolhuis, C.~Dellago, and D.~Chandler,
\newblock Reaction coordinates of biomolecular isomerization,
\newblock Proc. Natl. Acad. Sci. USA {\bf 97}, 5877  (2000).

\bibitem{McGibbon17}
R.~T. McGibbon, B.~E. Husic, and V.~S. Pande,
\newblock Identification of simple reaction coordinates from complex dynamics,
\newblock J. Chem. Phys. {\bf 146}, 044109 (2017).

\bibitem{Sittel18}
F.~Sittel and G.~Stock,
\newblock Perspective: Identification of collective coordinates and metastable
  states of protein dynamics,
\newblock J. Chem. Phys. {\bf 149}, 150901 (2018).

\bibitem{Fleetwood21}
O.~Fleetwood, J.~Carlsson, and L.~Delemotte,
\newblock Identification of ligand-specific g protein-coupled receptor states
  and prediction of downstream efficacy via data-driven modeling,
\newblock eLife {\bf 10}, e60715 (2021).

\bibitem{Diez22}
G.~Diez, D.~Nagel, and G.~Stock,
\newblock Correlation-based feature selection to identify functional dynamics
  in proteins,
\newblock J. Chem. Theory Comput. {\bf 18}, 5079 – 5088 (2022).

\bibitem{Wang21}
D.~Wang and P.~Tiwary,
\newblock State predictive information bottleneck,
\newblock J. Chem. Phys. {\bf 154}, 134111 (2021).

\bibitem{Glielmo21}
A.~Glielmo, B.~E. Husic, A.~Rodriguez, C.~Clementi, F.~No{\'e}, and A.~Laio,
\newblock Unsupervised learning methods for molecular simulation data,
\newblock Chem. Rev. {\bf 121}, 9722 (2021).

\bibitem{Sittel17}
F.~Sittel, T.~Filk, and G.~Stock,
\newblock Principal component analysis on a torus: Theory and application to
  protein dynamics,
\newblock J. Chem. Phys. {\bf 147}, 244101 (2017).

\bibitem{Perez-Hernandez13}
G.~Perez-Hernandez, F.~Paul, T.~Giorgino, G.~De~Fabritiis, and F.~No{\'e},
\newblock Identification of slow molecular order parameters for {Markov} model
  construction,
\newblock J. Chem. Phys. {\bf 139}, 015102 (2013).

\bibitem{Rohrdanz13}
M.~A. Rohrdanz, W.~Zheng, and C.~Clementi,
\newblock Discovering mountain passes via torchlight: Methods for the
  definition of reaction coordinates and pathways in complex macromolecular
  reactions,
\newblock Annu. Rev. Phys. Chem. {\bf 64}, 295 (2013).

\bibitem{Mardt18}
A.~Mardt, L.~Pasquali, H.~Wu, and F.~No{\'e},
\newblock {VAMPnets for deep learning of molecular kinetics},
\newblock Nat. Comm. {\bf 9}, 5 (2018).

\bibitem{Chen19}
W.~Chen, H.~Sidky, and A.~L. Ferguson,
\newblock Nonlinear discovery of slow molecular modes using state-free
  reversible {VAMPnets},
\newblock J. Chem. Phys. {\bf 150}, 214114 (2019).

\bibitem{Lemke19}
T.~Lemke and C.~Peter,
\newblock Encodermap: Dimensionality reduction and generation of molecule
  conformations,
\newblock J. Chem. Theory Comput. {\bf 15}, 1209 (2019).

\bibitem{Lemke19b}
T.~Lemke, A.~Berg, A.~Jain, and C.~Peter,
\newblock Encodermap ({II}): Visualizing important molecular motions with
  improved generation of protein conformations,
\newblock J. Chem. Inf. Model. {\bf 59}, 4550 (2019).

\bibitem{Belkacemi21}
Z.~Belkacemi, P.~Gkeka, T.~Leli{\`e}vre, and G.~Stoltz,
\newblock Chasing collective variables using autoencoders and biased
  trajectories,
\newblock J. Chem. Theory Comput. {\bf 18}, 59 (2021).

\bibitem{Hornik89}
K.~Hornik, M.~Stinchcombe, and H.~White,
\newblock Multilayer feedforward networks are universal approximators,
\newblock Neural Netw. {\bf 2}, 359 (1989).

\bibitem{Kingma13}
D.~P. Kingma,
\newblock Auto-encoding variational bayes,
\newblock arXiv:1312.6114  (2013).

\bibitem{Doersch16}
C.~Doersch,
\newblock Tutorial on variational autoencoders,
\newblock arXiv:1606.05908  (2016).

\bibitem{Ribeiro18}
J.~M.~L. Ribeiro, P.~Bravo, Y.~Wang, and P.~Tiwary,
\newblock Reweighted autoencoded variational bayes for enhanced sampling
  (rave),
\newblock J. Chem. Phys. {\bf 149}, 072301 (2018).

\bibitem{Varolgunecs20}
Y.~B. Varolg{\"u}ne{\c{s}}, T.~Bereau, and J.~F. Rudzinski,
\newblock Interpretable embeddings from molecular simulations using gaussian
  mixture variational autoencoders,
\newblock Mach. Learn.: Sci. Technol. {\bf 1}, 015012 (2020).

\bibitem{Tian21}
H.~Tian, X.~Jiang, F.~Trozzi, S.~Xiao, E.~C. Larson, and P.~Tao,
\newblock Explore protein conformational space with variational autoencoder,
\newblock Front. Mol. Biosci. {\bf 8}, 781635 (2021).

\bibitem{Tomczak18}
J.~Tomczak and M.~Welling,
\newblock {VAE} with a {V}amp{P}rior,
\newblock in {\em Int. Conf. Artif. Intell. Stat.}, pages 1214--1223, PMLR,
  2018.

\bibitem{Chung15}
J.~Chung, K.~Kastner, L.~Dinh, K.~Goel, A.~C. Courville, and Y.~Bengio,
\newblock A recurrent latent variable model for sequential data,
\newblock Adv. Neural Inf. Process. Syst. {\bf 28} (2015).

\bibitem{Girin20}
L.~Girin, S.~Leglaive, X.~Bie, J.~Diard, T.~Hueber, and X.~Alameda-Pineda,
\newblock Dynamical variational autoencoders: A comprehensive review,
\newblock arXiv:2008.12595  (2020).

\bibitem{Hasan21}
A.~Hasan, J.~M. Pereira, S.~Farsiu, and V.~Tarokh,
\newblock Identifying latent stochastic differential equations,
\newblock IEEE Trans. Signal Process. {\bf 70}, 89 (2021).

\bibitem{Wang24}
D.~Wang, Y.~Wang, L.~Evans, and P.~Tiwary,
\newblock From latent dynamics to meaningful representations,
\newblock J. Chem. Theory Comput. {\bf 20}, 3503 (2024).

\bibitem{Rabiner86}
L.~Rabiner and B.~Juang,
\newblock An introduction to hidden markov models,
\newblock IEEE ASSP Mag. {\bf 3}, 4 (1986).

\bibitem{Noe13}
F.~No{\'e}, H.~Wu, J.-H. Prinz, and N.~Plattner,
\newblock Projected and hidden markov models for calculating kinetics and
  metastable states of complex molecules,
\newblock J. Chem. Phys. {\bf 139} (2013).

\bibitem{Rasmussen03}
C.~E. Rasmussen,
\newblock Gaussian processes in machine learning,
\newblock in {\em Summer school on machine learning}, pages 63--71, Springer,
  2003.

\bibitem{Noe20}
F.~Noe, A.~Tkatchenko, K.-R. M\"uller, and C.~Clementi,
\newblock Machine learning for molecular simulation,
\newblock Annu. Rev. Phys. Chem. {\bf 71}, 361 (2020).

\bibitem{Higgins17}
I.~Higgins, L.~Matthey, A.~Pal, C.~P. Burgess, X.~Glorot, M.~M. Botvinick,
  S.~Mohamed, and A.~Lerchner,
\newblock $\beta$-{VAE}: {Learning} basic visual concepts with a constrained
  variational framework,
\newblock in {\em 5th Int. Conf. Learn. Represent. (ICLR)}, 2017.

\bibitem{Casale18}
F.~P. Casale, A.~Dalca, L.~Saglietti, J.~Listgarten, and N.~Fusi,
\newblock Gaussian process prior variational autoencoders,
\newblock Adv. Neural Inf. Process. Syst. {\bf 31} (2018).

\bibitem{Jazbec21}
M.~Jazbec, M.~Ashman, V.~Fortuin, M.~Pearce, S.~Mandt, and G.~R{\"a}tsch,
\newblock Scalable gaussian process variational autoencoders,
\newblock in {\em Int. Conf. Artif. Intell. Stat.}, pages 3511--3519, PMLR,
  2021.

\bibitem{Tian24}
T.~Tian, J.~Zhang, X.~Lin, Z.-L. Wang, and H.~Hakonarson,
\newblock Dependency-aware deep generative models for multitasking analysis of
  spatial omics data,
\newblock Nat. Methods {\bf 21}, 1501 (2024).

\bibitem{Titsias09}
M.~Titsias,
\newblock Variational learning of inducing variables in sparse gaussian
  processes,
\newblock in {\em Int. Conf. Artif. Intell. Stat.}, pages 567--574, PMLR, 2009.

\bibitem{Hensman13}
J.~Hensman, N.~Fusi, and N.~D. Lawrence,
\newblock Gaussian processes for big data,
\newblock arXiv:1309.6835  (2013).

\bibitem{Deringer21}
V.~L. Deringer, A.~P. Bart{\'o}k, N.~Bernstein, D.~M. Wilkins, M.~Ceriotti, and
  G.~Cs{\'a}nyi,
\newblock Gaussian process regression for materials and molecules,
\newblock Chem. Rev. {\bf 121}, 10073 (2021).

\bibitem{Mohr22}
B.~Mohr, K.~Shmilovich, I.~S. Kleinw{\"a}chter, D.~Schneider, A.~L. Ferguson,
  and T.~Bereau,
\newblock Data-driven discovery of cardiolipin-selective small molecules by
  computational active learning,
\newblock Chem. Sci. {\bf 13}, 4498 (2022).

\bibitem{Rasmussen06}
C.~E. Rasmussen and C.~K.~I. Williams,
\newblock {\em Gaussian Processes for Machine Learning}, volume~2,
\newblock MIT press Cambridge, MA, 2006.

\bibitem{Pearce20}
M.~Pearce,
\newblock The {G}aussian process prior {VAE} for interpretable latent dynamics
  from pixels,
\newblock in {\em Symposium on advances in approximate {B}ayesian inference},
  pages 1--12, PMLR, 2020.

\bibitem{Quinonero05}
J.~Quinonero-Candela and C.~E. Rasmussen,
\newblock A unifying view of sparse approximate gaussian process regression,
\newblock J. Mach. Learn. Res. {\bf 6}, 1939 (2005).

\bibitem{Jain14}
A.~Jain and G.~Stock,
\newblock Hierarchical folding free energy landscape of {HP35} revealed by most
  probable path clustering,
\newblock J. Phys. Chem. B {\bf 118}, 7750  (2014).

\bibitem{Nagel19}
D.~Nagel, A.~Weber, B.~Lickert, and G.~Stock,
\newblock Dynamical coring of {Markov} state models,
\newblock J. Chem. Phys. {\bf 150}, 094111 (2019).

\bibitem{Nagel23a}
D.~Nagel and G.~Stock,
\newblock msmhelper: A {Python package for Markov} state modeling of protein
  dynamics,
\newblock J. Open Source Softw. {\bf 8}, 5339 (2023).

\bibitem{Ernst17}
M.~Ernst, S.~Wolf, and G.~Stock,
\newblock Identification and validation of reaction coordinates describing
  protein functional motion: Hierarchical dynamics of {T4 Lysozyme},
\newblock J. Chem. Theory Comput. {\bf 13}, 5076  (2017).

\bibitem{Post22a}
M.~Post, B.~Lickert, G.~Diez, S.~Wolf, and G.~Stock,
\newblock Cooperative protein allosteric transition mediated by a fluctuating
  transmission network,
\newblock J. Mol. Bio. {\bf 434}, 167679 (2022).

\bibitem{Nagel24}
D.~Nagel, G.~Diez, and G.~Stock,
\newblock Accurate estimation of the normalized mutual information of
  multidimensional data,
\newblock J. Chem. Phys. {\bf 161}, 054108 (2024).

\bibitem{Ernst15}
M.~Ernst, F.~Sittel, and G.~Stock,
\newblock Contact- and distance-based principal component analysis of protein
  dynamics,
\newblock J. Chem. Phys. {\bf 143}, 244114 (2015).

\bibitem{Nagel23b}
D.~Nagel, S.~Sartore, and G.~Stock,
\newblock Toward a benchmark for {Markov} state models: The folding of {HP35},
\newblock J. Phys. Chem. Lett. {\bf 14}, 6956–6967 (2023).

\bibitem{Paszke19}
A.~Paszke,
\newblock Pytorch: An imperative style, high-performance deep learning library,
\newblock arXiv:1912.01703  (2019).

\end{thebibliography}


\begin{thebibliography}{1}

\bibitem{Rasmussen06}
C.~E. Rasmussen and C.~K.~I. Williams,
\newblock {\em Gaussian Processes for Machine Learning}, volume~2,
\newblock MIT press Cambridge, MA, 2006.

\bibitem{Truong20}
C.~Truong, L.~Oudre, and N.~Vayatis,
\newblock Selective review of offline change point detection methods,
\newblock Signal Process. {\bf 167}, 107299 (2020).

\bibitem{Killick12}
R.~Killick, P.~Fearnhead, and I.~A. Eckley,
\newblock Optimal detection of changepoints with a linear computational cost,
\newblock J. Am. Stat. Assoc. {\bf 107}, 1590 (2012).

\bibitem{Paszke19}
A.~Paszke,
\newblock Pytorch: An imperative style, high-performance deep learning library,
\newblock arXiv:1912.01703  (2019).

\bibitem{Loshchilov17}
I.~Loshchilov and F.~Hutter,
\newblock Decoupled weight decay regularization,
\newblock arXiv:1711.05101  (2017).

\bibitem{Jain12}
A.~Jain and G.~Stock,
\newblock Identifying metastable states of folding proteins,
\newblock J. Chem. Theory Comput. {\bf 8}, 3810  (2012).

\bibitem{Diez22}
G.~Diez, D.~Nagel, and G.~Stock,
\newblock Correlation-based feature selection to identify functional dynamics
  in proteins,
\newblock J. Chem. Theory Comput. {\bf 18}, 5079 – 5088 (2022).

\bibitem{Nagel24}
D.~Nagel, G.~Diez, and G.~Stock,
\newblock Accurate estimation of the normalized mutual information of
  multidimensional data,
\newblock J. Chem. Phys. {\bf 161}, 054108 (2024).

\bibitem{Post22a}
M.~Post, B.~Lickert, G.~Diez, S.~Wolf, and G.~Stock,
\newblock Cooperative protein allosteric transition mediated by a fluctuating
  transmission network,
\newblock J. Mol. Bio. {\bf 434}, 167679 (2022).

\end{thebibliography}

\end{document}


\title{Supplementary Material for:\\
Recovering Hidden Degrees of Freedom Using Gaussian Processes}

\author{Georg Diez}
\email{georg.diez@physik.uni-freiburg.de}
\author{Nele Dethloff}
\author{Gerhard Stock}
\email{stock@physik.uni-freiburg.de}
\affiliation{Biomolecular Dynamics, Institute of Physics, University of Freiburg, 79104 Freiburg, Germany}
\date{\today}


\vspace*{-2cm} 
{\centering
\large Supplementary Material for: \\
\textbf{Recovering Hidden Degrees of Freedom Using Gaussian Processes} \par
\vspace{2mm}
\normalsize
Georg Diez,\textsuperscript{1,a}
Nele Dethloff,\textsuperscript{1}
Gerhard Stock\textsuperscript{1,b} \\[1ex]
\textsuperscript{1}Institute of Physics, University of Freiburg, Freiburg, Germany \\[2ex]
\texttt{\textsuperscript{a}georg.diez@physik.uni-freiburg.de}; \texttt{\textsuperscript{b}stock@physik.uni-freiburg.de}
\par
\vspace{2ex}
{\centering \today \par}
}

\baselineskip5.4mm

\section{Matérn Kernel}
The Matérn kernel\cite{Rasmussen06} allows to independently control 
two different aspects of memory through the parameters $\nu$ and $\ell$.
The smoothness parameter $\nu$ determines how memory is structured 
by controlling the order of the underlying Markov process, while the 
length scale $\ell$ governs how quickly/slowly this memory decays over time.

We consider three cases, $\nu = 1/2, 3/2, \infty$:
\begin{align}
k_{\nu=1/2}(t, t'; \ell) &= \exp \left( -\frac{|t - t'|}{\ell} \right),\\
k_{\nu=3/2}(t, t'; \ell) &= \left(1 + \frac{\sqrt{3}|t - t'|}{\ell}\right) \exp \left( -\frac{\sqrt{3}|t - t'|}{\ell} \right),\\
\lim_{\nu \to \infty} k_\nu (t, t'; \ell) &= \exp \left( -\frac{|t - t'|^2}{2\ell^2} \right).
\end{align}
For $\nu = 1/2$, the kernel reduces to exponentially decaying memory corresponding
to a first-order Markov process (equivalent to the Ornstein-Uhlenbeck
covariance). Higher half-integer values ($\nu = 3/2, 5/2, ...$) yield higher-order
Markov processes with increasingly smooth realizations that retain more past
information. In the limit $\nu \rightarrow \infty$, the Matérn kernel converges 
to the RBF kernel, representing an infinitely smooth process with no Markovianity.

The length scale $\ell$ controls memory decay rate: large $\ell$ values provide 
slow decay and high predictive power, while small $\ell$ values yield fast decay 
and weak predictability.

\begin{figure}[H]
    \centering
    \includegraphics[width=0.8\linewidth]{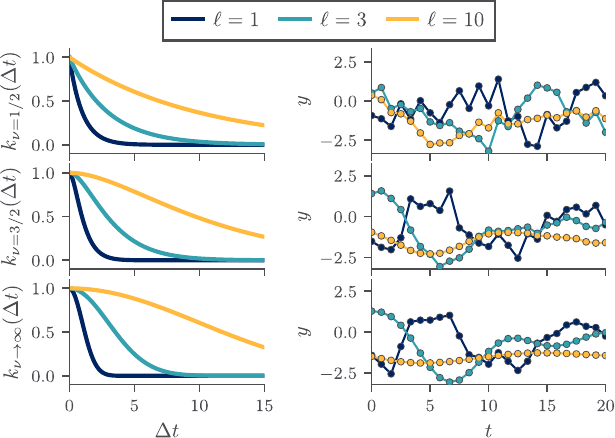}
    \caption{\baselineskip4mm
        Matérn kernel properties and Gaussian process realizations. 
        We define $\Delta t = |t-t^\prime|$.
        Left: Kernel correlation functions $k_\nu(\Delta t; \ell)$ for 
        different length scales $\ell \in \{1, 3, 10\}$ and smoothness 
        parameters $\nu \in \{1/2, 3/2, \infty \}$. 
        Right: Sample realizations from GP priors using the corresponding 
        Matérn kernels.
    }
    \label{si:fig:matern}
\end{figure}

Fig.~\ref{si:fig:matern} demonstrates how these parameters work together. 
The left panels reveal how varying $\ell \in \{1, 3, 10\}$ changes the 
width of memory decay, while different $\nu$ values $\in \{1/2, 3/2, \infty\}$ 
alter its shape. 
The GP samples on the right show the combined effect: higher $\nu$ values 
produce smoother paths regardless of the length scale, while larger $\ell$ 
values create slower, more predictable variations.

\section{Estimation of the Inducing Points}
\label{si:sec:inducingpoints}
As we rely on sparse approximations for the latent GP regression, we need
to estimate the inducing points $\bm{U} = [\bm{u}_1, \dots, \bm{u}_{n_u}]$. 
In order to obtain inducing points that are as representative as possible
of the system's dynamics, 
we employ a \textit{change point detection} algorithm.\cite{Truong20}
The Pruned Exact Linear Time\cite{Killick12} (PELT) algorithm effectively detects
significant changes in the time traces of the system by minimizing
the following cost function using dynamic programming:
\begin{align}
    F(t) = \min_{\tau < t} \left[F(\tau) + C(\mathbf{x}_{\tau + 1:t}) + \beta \right]
\end{align}
Here, $F(t)$ denotes the optimal partitioning up to time $t$, 
$\mathbf{x}_{1:T} = (\mathbf{x}_1, \dots, \mathbf{x_T})$ is the 
(multi-dimensional and ordered) trajectory,
$C(\mathbf{x}_{\tau + 1:t})$ a cost function measuring the homogeneity of
$\mathbf{x}$ within the segment $\tau+1$ till $t$, and $\beta$ is a penalty 
term preventing over-segmentation.
This approach automatically identifies time points when significant 
changes (i.e., conformational changes) occur. 

\clearpage
\section{Model Hyperparameters}
\label{si:sec:parameters}
For the numerical studies, we implemented the models using 
PyTorch\cite{Paszke19} with the following architecture and 
hyperparameters:

\begin{table}[h!]
\centering
\begin{tabular}{|p{6.5cm}|p{3.4cm}|p{3.1cm}|p{3.5cm}|}
\hline
\textbf{Parameter} & \textbf{Analytical Toy Model (GP-VAE)} & \textbf{Analytical Toy Model  (VAE)} & \textbf{T4L (GP-VAE)} \\
\hline
\multicolumn{4}{|c|}{\textbf{Neural Network Architecture}} \\
\hline
Hidden dimensions & 10-32-64-32-10 & 10-32-64-32-10 & 32-32 \\
\hline
\multicolumn{4}{|c|}{\textbf{Training Configuration}} \\
\hline
Optimizer & AdamW\cite{Loshchilov17} & AdamW\cite{Loshchilov17} & AdamW\cite{Loshchilov17}\\
Learning rate & $10^{-5}$ & $10^{-5}$ & $10^{-3}$ \\
Weight decay & $10^{-2}$ & $10^{-2}$ &$10^{-3}$ \\
Batch size & 5000$^*$ & 32 & 10000$^*$ \\
Training epochs & 100 & 100 & 100 \\
\hline
\multicolumn{4}{|c|}{\textbf{Model Hyperparameters}} \\
\hline
KL-divergence weight $\beta$ & 20 & 0.01$^{**}$ & 10 \\
Matérn kernel smoothness parameter $\nu$ & 3/2 & - & 3/2 \\
Matérn kernel length scale $l$ [frames]& $7.5 \cdot 10^4$ & - & $7 \cdot 10^4 \hat{=}7\,\mu$s\\
\hline
\end{tabular}
\caption{Model parameters and hyperparameters}
\label{tab:model_params}
\end{table}
For all experiments, we used the LeakyReLU functions, batch normalization, 
and a MinMax input scaler.

$^*$ For the GP-VAE applications, the batch size is chosen significantly 
larger than for traditional applications, since the GP regression 
quantities $\boldsymbol{\mu}$, $\mathbf{A}$ and $\boldsymbol{\Sigma}$ converge to the 
true values for large batch sizes.\\
$^{**}$ We tested $\beta$-values in the range of $10^{-3}-10^{3}$, but none yielded
results that successfully separated the two overlapping states in the toy model. 
The geometric structure of the $xy$-plane projection is best preserved with 
relatively small $\beta$-values, as larger values over-regularize the 
latent space.   

\section{Supplementary Results for the Toy Model}
\subsection{Inducing Points}
\begin{figure}[h]
    \centering
    \includegraphics[width=0.8\linewidth]{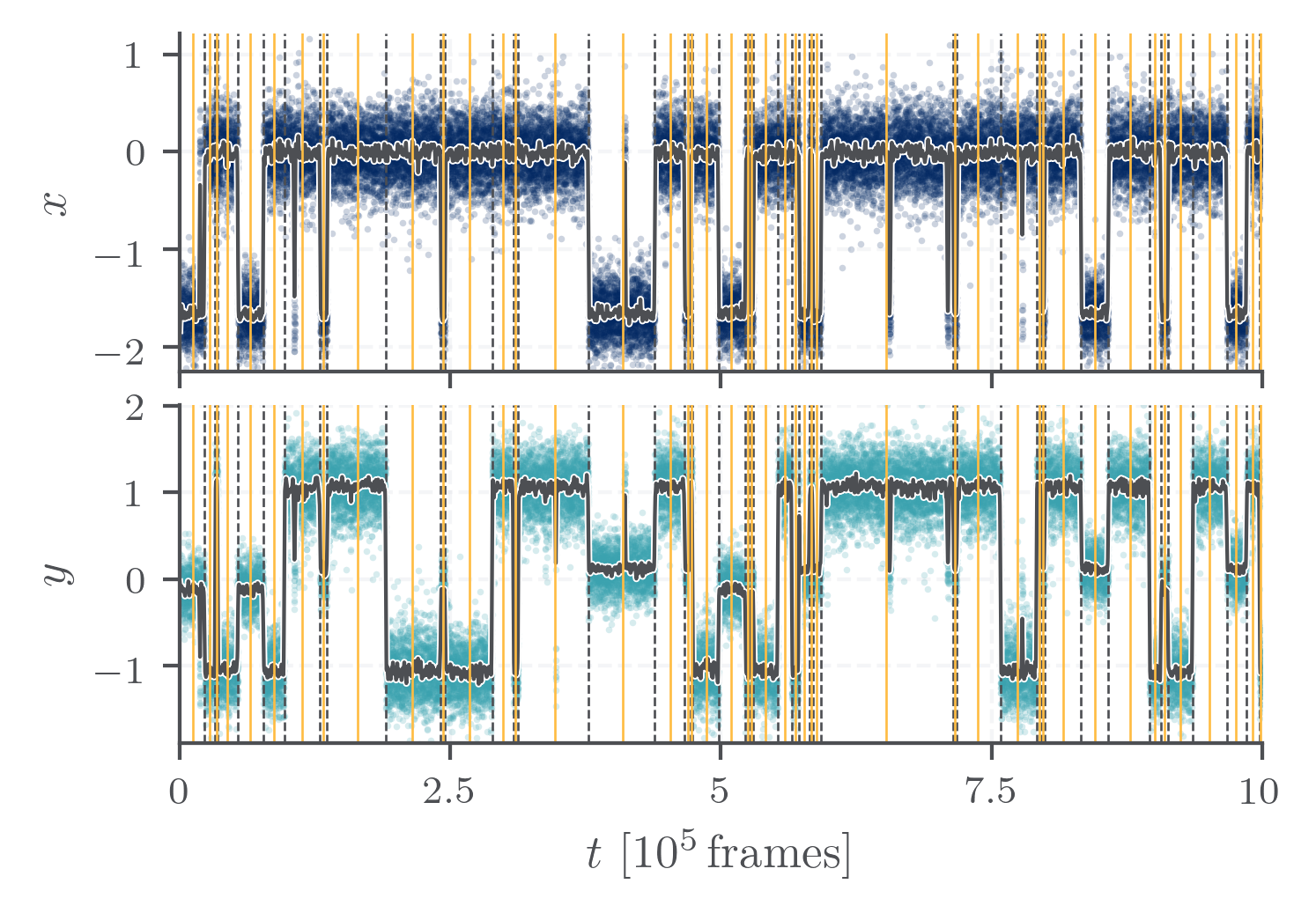}
    \caption{\baselineskip4mm
        Inducing Points identified by the PELT algorithm for the analytical
        toy model (yellow vertical lines), capturing major transitions 
        between the metastable states in the $xy$-plane.
        Additional time points were added at the midpoint of each segment
        in order to reflect the metastable conformation as well 
        (dashed gray lines).
    }
    \label{si:fig:inducing_points}
\end{figure}

For our specific analytical toy model, we analyze the trajectory restricted
to the $xy$-plane, where states 3 and 4 overlap.
This mimics the scenario in real MD simulations analysis, where important degrees 
of freedom may be hidden or overlooked in post-simulation analysis.
The PELT algorithm identified 45 change points
shown as yellow vertical lines in the time traces in
Fig.~\ref{si:fig:inducing_points}. 
To represent the metastable states as well, we added a center point in the 
middle of each two pairs of inducing points.
This results in a total of $n_u=89$ inducing points.

\subsection{Vanilla VAE Embedding}
To investigate and isolate the impact of the GP prior, we perform a controlled
comparison between our GP-VAE approach and a vanilla VAE baseline.
The vanilla VAE employs identical network architecture and training
parameters, differing only in the KL-divergence weight $\beta$ and 
the batch-size (see Tab.~\ref{tab:model_params}).
This analogousness ensures that observed differences in the learned embedding(s)
can be attributed specifically to the temporal correlation structure imposed
by the GP prior.

\begin{figure}[H]
    \centering
    \includegraphics[width=0.8\linewidth]{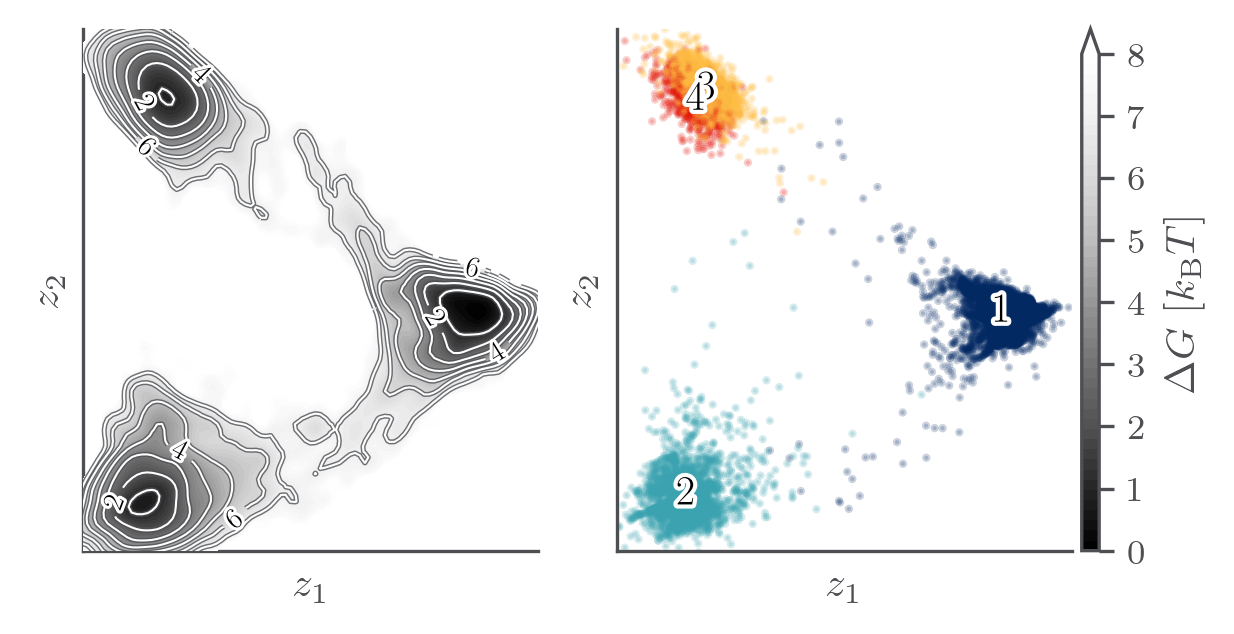}
    \caption{\baselineskip4mm
        Vanilla VAE embedding of the analytical toy model using only xy-plane coordinates 
        with $\beta = 0.01$. 
        The left panel shows the vanilla VAE embedding of the $xy$-coordinates,
        while the right panel displays the same embedding colored according to
        the original four-state clustering from the full three-dimensional
        space.
        The standard VAE fails to recover the hidden dynamics from 
        the missing z-dimension, showing the same overlapping of states 3 and 4 
        as observed in the $xy$-plane projection.
        State labels $1$-$4$ mark the centroid positions of each respective
        state cluster.
    }
    \label{SI:fig:vae_embedding}
\end{figure}

We tested a wide range of hyperparameters (different $\beta$-values, batch sizes
and learning rates) and the best looking resulting embedding is shown in 
Fig.~\ref{SI:fig:vae_embedding}.
Unlike the GP-VAE approach, the vanilla VAE cannot leverage temporal correlations 
to separate these geometrically identical but dynamically distinct states, 
which is why the vanilla VAE basically learns the identity projection.
This demonstrates the necessity of the GP prior for recovering Markovian dynamics 
from incomplete spatial information.

\subsection{Most Probable Path Coarse Graining}
\begin{figure}[H]
    \centering
    \includegraphics[width=0.9\linewidth]{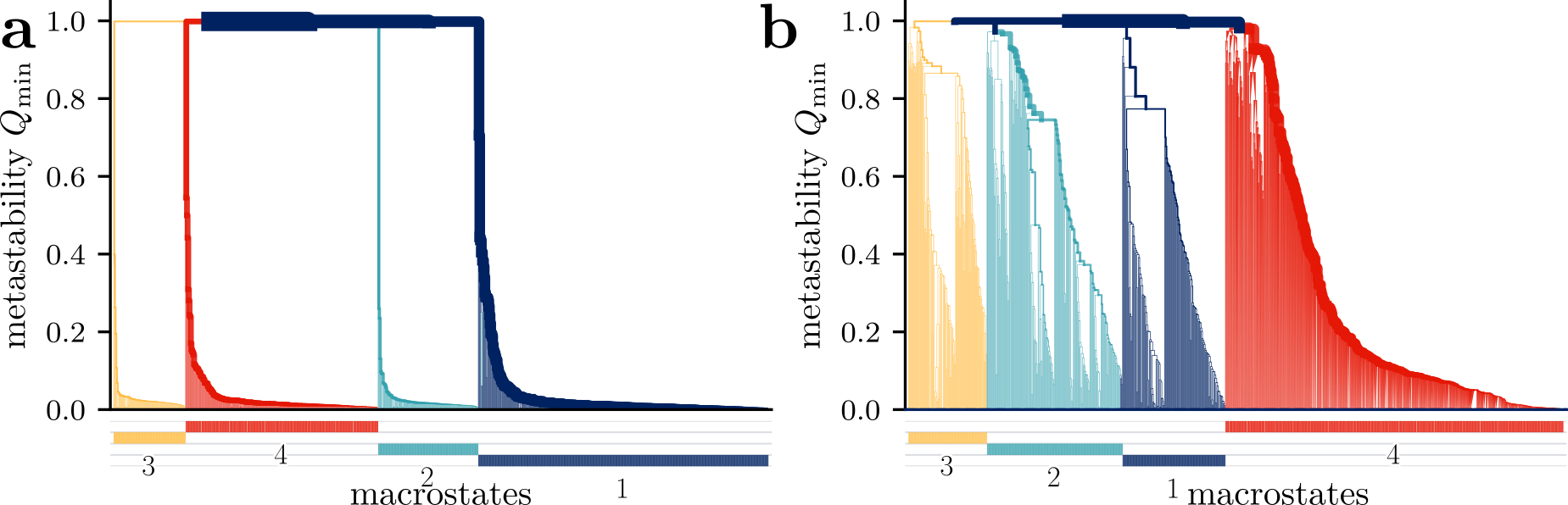}
    \caption{\baselineskip4mm
        Dendrogram illustrating the hierarchical lumping of the microstates into
        macrostates based on their metastability and transition probabilities
        for the toy model.\cite{Jain12} 
        The $y$-axis represents the minimum metastability $Q_\text{min}$,
        which determines the state merging.
        Once $Q_\text{min}$ surpasses the 
        self transition probability of a state $i$ $T_{i\rightarrow i}(\tau)$,
        the state $i$ is merged into another (branch of) microstate(s) $j$ 
        which features the highest transition probability 
        $\max_j T_{i \rightarrow j}(\tau)$.
        This process progressively combines microstates into larger 
        macrostates as $Q_\text{min}$ increases from 0 to 1.
        (a) Hierarchical clustering of microstates in the full
        three-dimensional space. 
        (b) Corresponding lumping in the GP-VAE embedding space.
    }
    \label{si:fig:dendrograms}
\end{figure}

\section{Supplementary Results for T4 Lysozyme}

\subsection{T4L Inducing Points}
\begin{figure}[H]
    \centering
    \includegraphics[width=0.9\linewidth]{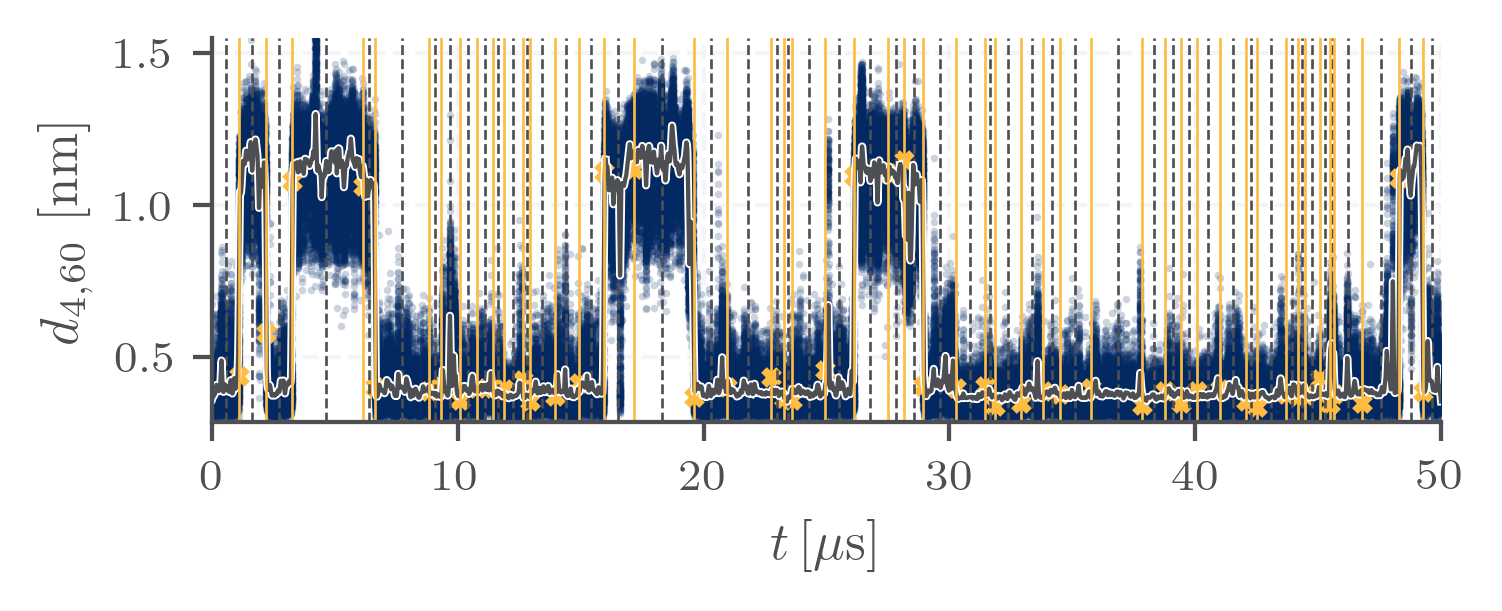}
    \caption{\baselineskip4mm
        Inducing Points determined by the PELT algorithm employed on the 
        locking distance $d_{4,60}$ which faithfully captures whether
        the system is in either the open or closed conformation.
        Additional time points were added at the midpoint of each segment
        in order to reflect the metastable conformation as well 
        (dashed gray lines).
    }
    \label{si:fig:inducing_points_t4l}
\end{figure}

For determining a suitable set of inducing points for T4 lysozyme, we applied the above described
PELT algorithm to the locking distance $d_{4,60}$. 
The resulting 51 change points are shown in Fig.~\ref{si:fig:inducing_points_t4l}.
Additionally, we added a center point between each pair of inducing points  
to represent metastable conformations, resulting in a total of $n_u=101$
inducing points.

\subsection{MoSAIC analysis}
\begin{figure}[H]
    \centering
    \includegraphics[width=0.5\linewidth]{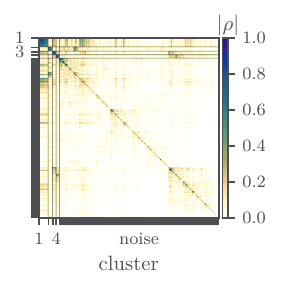}
    \caption{\baselineskip4mm
    Block-diagonalized correlation matrix of the 556 identified contact distances
    of T4L using MoSAIC with the CPM model and $\gamma=0.5$.}
    \label{fig:mosaic_mat}
\end{figure}

Requiring a minimum contact occupancy of 1\,\%, we identified from the 50\,$\mu$s-long MD trajectory 556 contact distances, for which
we computed the absolute linear correlation matrix
\begin{align}
    |\rho| = \frac{ \langle \delta x_\alpha \delta x_\beta \rangle}{\langle \delta x_\alpha^2 \rangle^{1/2}\langle \delta x_\beta^2 \rangle^{1/2}},
\end{align}
where $\delta x_\alpha = x_\alpha - \langle x_\alpha \rangle$ and $\langle \dots \rangle$
denotes the time average over the MD data.
Employing MoSAIC\cite{Diez22} using the constant Potts model with a resolution parameter of $\gamma=0.5$ on the resulting correlation matrix results in the 
block-diagonalized correlation matrix shown in Fig.~\ref{fig:mosaic_mat}.
The distances contained in each cluster are listed in Tab.~\ref{SI:tab:clusters}
and structurally in Fig.~4 in the main paper.
We note that the resulting number of $32$ correlated
    contacts in cluster C1 agrees with the results
    of Ref.~\citenum{Nagel24}, but differs to the results of 
    Ref.~\citenum{Post22a}, that mistakenly included hydrogen atoms in
    the contact definition, which resulted in $\sim 85$ correlated
    contacts.

\begin{table}[h!]
    \centering
    \caption[Coordinates in MoSAIC cluster 1]{
        Inter-residue distances and first side-chain dihedral angles within
        clusters found by MoSAIC, sorted by their average
        correlation within the cluster.
        \label{SI:tab:clusters}
    }
    \begin{tabular}{cl}
    \textbf{Cluster} & \textbf{Coordinates} \\
    \hline
    1 & $d_{4,60}$, $d_{4,63}$, $d_{4,13}$, $d_{4,29}$, $d_{4,72}$, $d_{22,137}$, $d_{4,64}$, $d_{20,142}$, $d_{8,67}$, $d_{8,68}$\\
     & $d_{22,141}$, $d_{21,141}$, $d_{2,64}$, $d_{7,71}$, $d_{1,64}$, $d_{4,71}$, $d_{30,145}$, $d_{21,142}$, $d_{5,60}$\\
     & $d_{7,12}$, $d_{8,64}$, $d_{20,145}$, $d_{4,68}$, $d_{8,13}$, $d_{3,67}$, $d_{5,64}$, $d_{24,105}$, $d_{8,12}$, \\
     & $d_{29,64}$, $d_{11,20}$, $d_{2,67}$, $d_{11,30}$ \\
    2 & $d_{10,101}$, $d_{6,98}$, $d_{6,97}$, $d_{9,161}$, $d_{6,94}$, $d_{9,160}$, $d_{10,149}$, $d_{10,105}$, $d_{9,158}$, \\
     &  $d_{10,145}$, $d_{6,152}$, $d_{9,148}$, $d_{6,101}$, $d_{3,100}$ \\
    3 & $d_{20,24}$, $d_{20,25}$, $d_{18,22}$, $d_{22,26}$, $d_{14,20}$, $d_{14,21}$, $d_{22,30}$, $d_{20,32}$, $d_{20,26}$ \\
     & $d_{11,22}$ \\
    4 & $d_{36,42}$, $d_{25,34}$, $d_{36,45}$, $d_{24,34}$, $d_{34,38}$, $d_{34,41}$, $d_{34,42}$, $d_{37,41}$, $d_{23,34}$ \\
     & $d_{35,45}$ \\
    \end{tabular}
\end{table}

\subsection{Structural Analysis of the GP states}
\begin{figure}[b]
    \centering
    \includegraphics[width=0.4\linewidth]{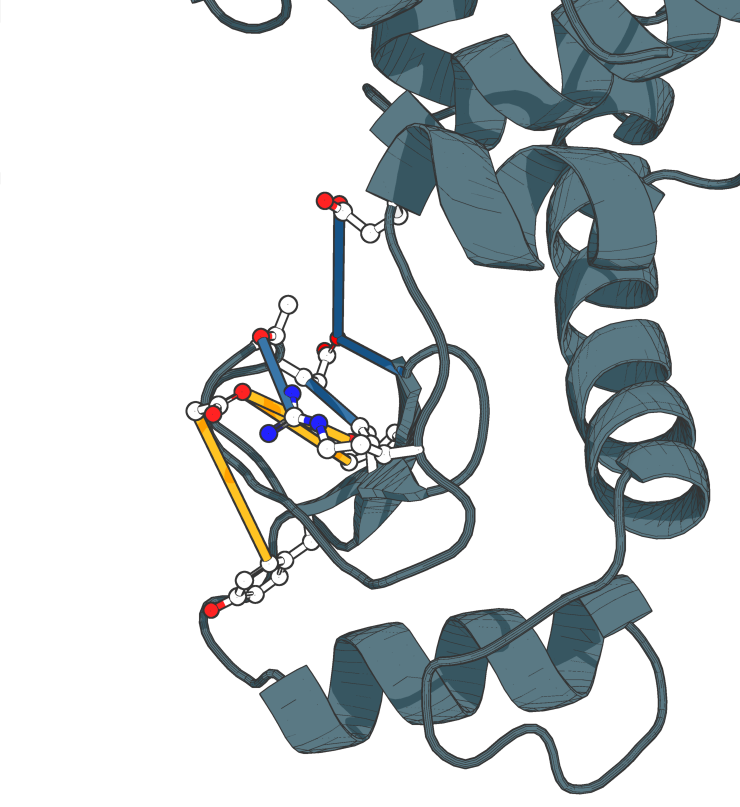}
    \hspace{0.5cm}
    \includegraphics[width=0.4\linewidth]{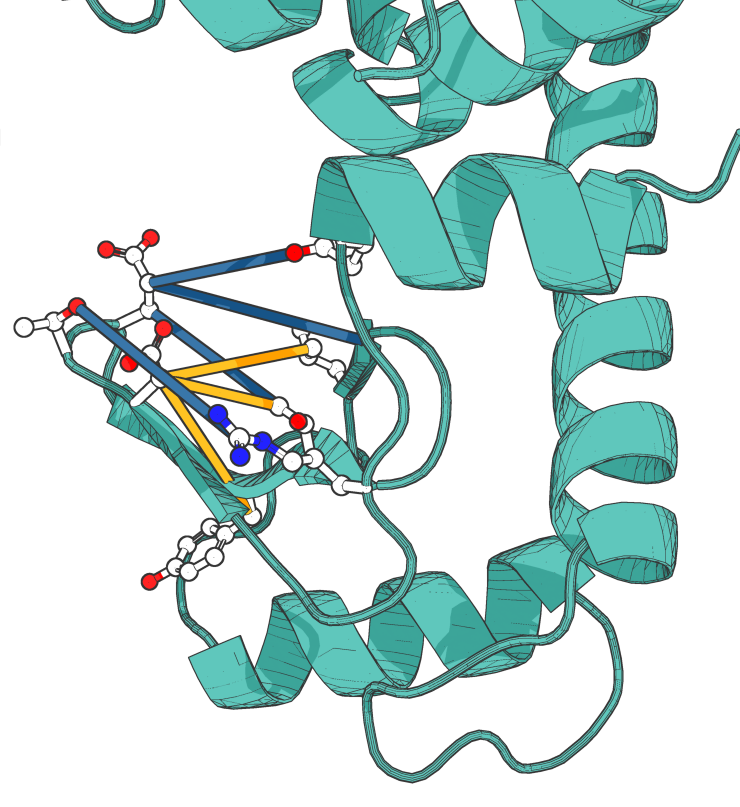}
    \caption{\baselineskip4mm
        Structural comparison of the two GP states in T4 lysozyme. 
        Left: State $1_\text{GP}$ showing the conformational arrangement 
        that enables the open$\leftrightarrow$closed motion, with 
        motion-enabling contacts formed (blue distances) and
        motion-restricting contacts unformed in the jaw-lip region. 
        Right: State $2_\text{GP}$ displaying the constrained conformation w
        here motion-restricting contacts are formed (yellow regions), 
        preventing global conformational transitions. 
        The structural differences in the jaw-lip region
        demonstrate how local contact rearrangements control the protein's
        functional dynamics.
    }
    \label{si:fig:structure}
\end{figure}
The structural analysis of the two open GP states---$1_\text{GP}$ and 
$2_\text{GP}$---in T4 lysozyme reveals distinct conformational differences 
that have important functional implications for the
T4L's open$\leftrightarrow$closed dynamics.
From each state we extract the structure with the lowest free energy
and show it in Fig.~\ref{si:fig:structure}---on the left $1_\text{GP}$
allowing open$\leftrightarrow$closed transition and on the right 
$2_\text{GP}$ which is the dynamically isolated state.

\begin{figure}[H]
    \centering
    \includegraphics[width=0.8\linewidth]{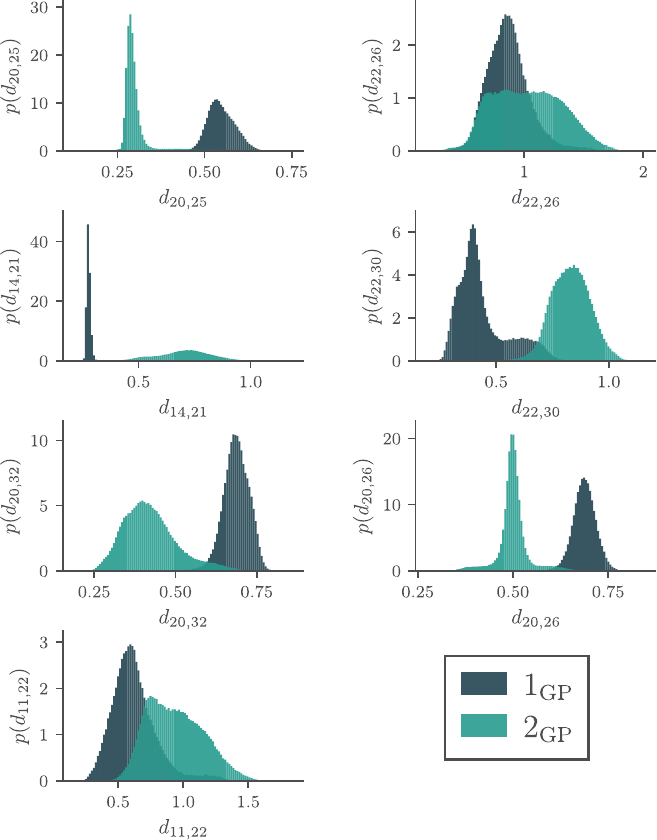}
    \caption{\baselineskip4mm
        Probability distributions of seven key inter-residue distances 
        characterizing the two GP states in T4L. The distributions 
        show the structural differences between state $1_\text{GP}$ 
        (allowing open$\leftrightarrow$closed dynamics) and state 
        $2_\text{GP}$ (dynamically isolated) along critical distances 
        that control the open$\leftrightarrow$closed transition.
    }
    \label{si:fig:hists}
\end{figure}

The probability distribution of both states along seven key distances
(we excluded three others due to their vanishing eigenvector contribution
to $x_1^{(3)}$) are shown in Fig.~\ref{si:fig:hists}
We can categorize the seven distances into two functional groups:
distances $d_{20,25}$, $d_{20,23}$ and $d_{20,26}$ (shown in yellow)
form contacts in $2_\text{GP}$ while remaining extended in $1_\text{GP}$,
and distances $d_{22,26}$, $d_{14,21}$, $d_{22,30}$ and $d_{11,22}$
(displayed in blue) form contacts in $1_\text{GP}$, while these
contacts are mostly broken in the dynamically isolated state $2_\text{GP}$.

The key structural difference between the two states lies in a
mechanism that governs the mobility in the jaw region:
in state $1_\text{GP}$, the blue contacts that connect the very  
outer end of the lip with the $\beta$-sheets and the 
$\alpha_1\beta$-loop, are formed.
Consequently, the motion of the lip is effectively restricted,  
creating a local conformation in the lower jaw that facilitates
interdomain contacts between the N- and C-domains across the 
mouth region and pre-organizes the jaw for the global
open$\leftrightarrow$closed motion.

In contrast, state $2_\text{GP}$ allows for increased flexibility 
in the lip: contacts formed by the yellow distances attach further 
inside on the lip, while the stabilizing blue contacts in $1_\text{GP}$
are not formed. This increased flexibility is functionally 
counterproductive since the open$\leftrightarrow$closed dynamics of 
T4L requires a highly coordinated and simultaneous switching of all 
relevant contacts in MoSAIC cluster 1, resulting in a cogwheel-like 
motion.\cite{Post22a}

\subsection{Full-dimensional Embedding of Cluster 1}
\begin{figure}[H]
    \centering
    \includegraphics[width=0.45\linewidth]{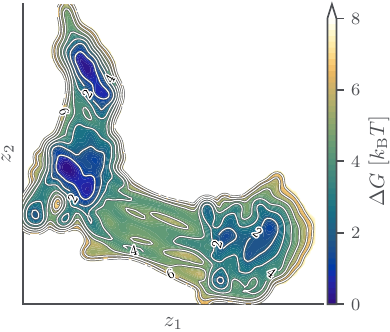}
    \caption{\baselineskip4mm
        GP-VAE embedding of all 32 contacts contained in MoSAIC
        cluster 1 (compare Tab.~\ref{SI:tab:clusters}). 
        We used the same hyperparameters as for the two dimensional model 
        (compare Tab.~\ref{tab:model_params}).
    }
    \label{SI:fig:t4l_c1_embedding}
\end{figure}

\bibliography{new,Bib/stock}